\documentclass[article,onecolumn,doublespacing]{IEEEtran}
\usepackage[T1]{fontenc}
\usepackage{graphicx,colordvi,psfrag}
\usepackage{amsmath,amssymb,algpseudocode}
\usepackage{algorithm}
\usepackage{algorithmicx}
\usepackage{calc,pstricks, pgf, xcolor}
\usepackage{bbding}
\usepackage{bbm}
\usepackage{dsfont}
\usepackage{stfloats}
\usepackage{hyperref}
\usepackage{xparse}
\usepackage{caption}
\usepackage{tikz}
\usepackage{forest}
\usetikzlibrary{calc,intersections}

\newcommand{\dist}{\mathrm{dist}}
\newcommand{\RR}{\mathbb{R}}

\newcommand{\E}{\mathbb{E}}
\newcommand{\EE}{\mathbb{E}}
\newcommand{\tr}{\top}
\newcommand{\Simplex}{\Delta_{m}}
\DeclareMathOperator*{\argmin}{\arg\!\min}
\DeclareMathOperator*{\argmax}{\arg\!\max}
\newcommand*\rfrac[1]{\frac1{1-#1}}
\newcommand{\vphi}{\varphi}
\newcommand{\CommaPunct}{\mathbin{\raisebox{0.5ex}{,}}}
\newcommand{\df}{{\, \stackrel{\mathrm{def}}{=}\, }}

\newtheorem{theorem}{Theorem}

\newtheorem{proposition}{Proposition}
\newtheorem{claim}{Claim}

\newtheorem{remark}{Remark}
\newtheorem{example}{Example}
\newtheorem{definition}{Definition}
\newtheorem{lemma}{Lemma}

\newtheorem{problem}{Problem}
 \newenvironment{proof}[1][Proof]{\noindent\textbf{#1.} }{\ \rule{0.5em}{0.5em}}

\def\m{\mathcal}

\title{Optimal Online Bookmaking for Binary Games}

\author{Alankrita Bhatt, %,~\IEEEmembership{Member,~IEEE}, 
	Or Ordentlich and Oron Sabag %~\IEEEmembership{Senior Member,~IEEE} 
\thanks{A. Bhatt is with the California Institute of Technology (\texttt{abhatt@caltech.edu}). O. Ordentlich and O. Sabag are with the Rachel and Selim Benin School of Computer Science and engineering,
Hebrew University of Jerusalem, Israel (\texttt{or.ordentlich@mail.huji.ac.il}, \texttt{oron.sabag@mail.huji.ac.il}). The work of AB was supported by the Center for the Mathematics of Information Postdoctoral Fellowship at Caltech. The work of OO was supported by the Israel Science 
Foundation (ISF),
grant No. 1641/21. The work of OS was supported by the Israel Science 
Foundation (ISF),
grant No. 1096/23. }
}

\begin{document}
\maketitle
\begin{abstract}
In online betting, the bookmaker can update the payoffs it offers on a particular event many times before the event takes place, and the updated payoffs may depend on the bets accumulated thus far. We study the problem of bookmaking with the goal of maximizing the return in the worst-case, with respect to the gamblers' behavior and the event's outcome. We formalize this problem as the \emph{Optimal Online Bookmaking game}, and provide the exact solution for the binary case. To this end, we develop the optimal bookmaking strategy, which relies on a new technique called bi-balancing trees, that assures that the house loss is the same for all \emph{decisive} betting sequences, where the gambler bets all its money on a single outcome in each round.
\end{abstract}

\section{Introduction and Main results}

Consider an experiment $I$ with $m\in\mathbb{N}$ possible outcomes, say $[m]=\{0,1,\ldots,m-1\}$. Bookmakers offer bets of the following form on the outcome of the experiment: A gambler may invest money in any of the $m$ outcomes. For any $1\$ $ invested in outcome $i$, the gambler receives $\gamma(i)\$ $ ($\gamma(i)\geq 1$) if the experiment's result is $I=i$, and $0\$ $ if the experiment's result is not~$i$.  Such bets are often referred to as ``horse races'' in the literature~\cite{CoverThomas}, though the underlying experiment need not be a horse race.

Let
\begin{align}
 \Gamma=\sum_{i=0}^{m-1}\frac{1}{\gamma(i)}   
\end{align}
be the \emph{overround} parameter associated with the offered bet, and
\begin{align}
r(i)=\frac{1}{\Gamma}\frac{1}{\gamma(i)},~i=0,\ldots,m-1, 
\label{eq:rorelation}
\end{align}
be the probability distribution the bookmaker ``assigns'' to the experiment. Assume further that the gambler invests $q(i)\$ $ in each outcome $i=0,\ldots,m-1$, where $\sum_{i=0}^{m-1} q(i)=1$, such that both $r=(r(0),\ldots,r(m-1))^\tr$ and $q=(q(0),\ldots,q(m-1))^\tr$ are vectors in the $(m-1)$-dimensional probability simplex
\begin{align}
\Simplex=\left\{p\in[0,1]^m~:~\sum_{i=0}^{m-1} p(i)=1 \right\}.
\end{align}
Throughout the paper, the word ``gambler'' need not refer to a single gambler, but to the accumulation of all gamblers placing their bets on the experiment's outcome through the bookmaker. From the bookmaker's perspective, the number of gamblers participating in the bet makes no difference, and it is only the total amount invested in each of the $m$ possible outcomes that matters.

The bookmaker has collected $1\$ $ from the gambler before the experiment, and needs to pay the gambler
\begin{align}
L=\sum_{i=1}^{m-1} \mathbbm{1}\{I=i\} q(i)\gamma(i) =\frac{1}{\Gamma}\sum_{i=0}^{m-1}\mathbbm{1}\{I=i\}f_i(r,q)
\end{align}
dollars after the experiment has taken place, where $f_i(r,q)=\frac{q(i)}{r(i)}$, $i=0,\ldots,m-1$.  The worst-case scenario, from the bookmaker's perspective, is that the experiment's outcome is $i^*=\argmax_i f_i(r,q)$, and in this case $L=\frac{1}{\Gamma}\|f(r,q)\|_{\infty}$, where $f(r,q)=(f_0(r,q),\ldots,f_{m-1}(r,q))^\tr$. Since $r,q\in\Simplex$ we clearly have that $\|f(r,q)\|_{\infty}=\max_{i}\frac{q(i)}{r(i)}\geq 1$ and this is attained with equality iff $r=q$. Thus, a risk-averse bookmaker, whose objective is maximal gain in the worst-case, should aim to choose $r$ in a way that will cause the gambler to distribute its budget as $q=r$. If it succeeds, the house will collect a gain of $1-\frac{1}{\Gamma}$ regardless of the experiment's outcome. Thus, when $\Gamma=1$ the bet is fair, and for $q=r$ both the house and the gambler get zero gain, but when $\Gamma>1$, the house has a positive gain and the gambler a negative gain, when $r=q$.\footnote{There is clearly no motivation for a single gambler to distribute its budget with $q=r$ in this case, as this will result in negative gain with probability~$1$. However, recall that our ``gambler'' is composed of many individual gamblers and $q$ represents their combined distribution, so some of them may obtain a positive gain even if $q=r$.} Note, however, that the bookmaker first declares $r$ and the gambler places its bets according to $q$ only afterwards. It is therefore impossible for the bookmaker to guarantee that $r=q$. On the other hand, if the gambler places its entire bet on the experiment's outcome, the house can gain only if its return is smaller than the placed bet, that is, $\gamma(i) \le 1$. Together with the odds requirement $\gamma(i)\ge1$, it simply means that unless the house odds are $\gamma(i)=1$ (achieved with $\Gamma = m, \gamma(i)\frac1{m}$), the house can lose under particular scenarios.

The game described above consists of a single round: before the experiment the bookmaker declares $\Gamma>0$ and $r\in\Simplex$ (which correspond to $\gamma(0),\ldots,\gamma(m-1)$ via~\eqref{eq:rorelation}) and those cannot be updated until the experiment takes place. The gambler then chooses $q\in\Simplex$ and distributes its budget accordingly. Today, however, the gambling process is typically much more dynamic. Sports betting is mostly performed through websites, which employ algorithms for updating the odds they offer as the event approaches, and sometimes even throughout the event~\cite{lopez2018understanding}. The algorithms computing the updated odds may rely on the bets accumulated on the event thus far. The update algorithms may further rely on other factors such as new information related to the event \cite{betting_markets_iclr} (for example, it may be revealed that a certain player is injured), but in this paper we ignore such additional opportunities, and we take a worst-case approach. 

We therefore consider the following setup. Assume the bet happens in $T$ rounds, and let $\Gamma\geq 1$ be a fixed overround parameter that remains constant throughout the $T$ rounds. In each round $t=1,\ldots,T$, the bookmaker declares $r_t\in\Simplex$, such that the return it offers for each outcome is $\gamma_t(i)=\frac{1}{\Gamma \cdot r_t(i)}$, and the gambler responds by investing a budget of $1\$ $ distributed on the $m$ possible outcomes according to $q_t$. The experiment happens at the end of the $T$th round. The bookmaker has collected $T\$ $ and needs to pay the gambler
\begin{align}
L=\frac{1}{\Gamma}\sum_{i=0}^{m-1}   \mathbbm{1}\{I=i\} \sum_{t=1}^T f_i(r_t,q_t) 
\end{align}
dollars after the experiment have taken place. The question we pursue is:

\medskip

\emph{``What is the largest gain the bookmaker can guarantee  regardless of the gambler's behavior and the experiment's outcome?''}

\medskip

We restrict attention to the case of a binary experiment ($m=2$) and provide an \emph{exact} answer: the largest gain that can be guaranteed is $T\left(1-\frac{1+T^{-1/2}}{\Gamma} \right)$. Recall that $(1-\frac{1}{\Gamma})$ is the gain in a single-round for the (non-causal) choice $r=q$. Our result shows that in an online bet with $T$ rounds we can come close to this gain, up to a (normalized) penalty of $T^{-1/2}/\Gamma$. In particular, the bookmaker can guarantee a positive gain whenever the overround parameter satisfies $\Gamma>1+T^{-1/2}$. 
We also provide the precise algorithm for updating $r_t$ based on $q_1,\ldots,q_{t-1}$ which attains this return, as well as an efficient algorithm that approximately attains this performance.

The problem of designing the update policy and computing the optimal gain is a sequential/online optimization problem. Very few problems in this family have been solved \emph{exactly} for any horizon $T$, see e.g.~\cite{cover_experts,rakhlin2016tutorial,gravin2016towards, cover1991universal, shtar1987universal}, and it is quite remarkable that this particular problem does admit an exact solution. Moreover, we find an efficient algorithm to compute the update policy. Furthermore, while the $O(T^{-1/2})$ behavior of the penalty term is common to online optimization problems, we could not find any of-the-shelf online optimization algorithm attaining this rate of convergence for our problem. The reason for this is that the loss functions $f_i(r,q)$ are unbounded. See more in Section~\ref{subsec:relatedwork}. 

\subsection{Mathematical formulation of the online bookmaking game}
\label{subsec:mathform}

We formalize our problem as a repeated vector-valued game with  $T$-rounds, dubbed \emph{the online bookmaking game}, with two players, the house/bookmaker and the gambler. The game takes place in $T$ rounds, where in each round the house chooses a point in the simplex, which represents the returns it offers for each one of the $m$ outcomes of the event gambled on, and then the gambler chooses a point in the simplex, which represents how it distributes its betting money on the $m$ outcomes. Define:
\begin{itemize}
\item \textbf{Loss vector}:  The loss vector is a mapping $f:\Simplex\times \Simplex\to \RR_+^{m}$. In particular,
\begin{align}
f(r,q)=(f_0(r,q),\ldots,f_{m-1}(r,q))^\tr,    
\end{align}
where
\begin{align}\label{eq:def_fi_intro}
f_i(r,q)=\frac{q(i)}{r(i)},~~~i=0,\ldots,m-1.
\end{align}
\item \textbf{House Strategy/Algorithm}: A strategy for the first player (the house) is a set of $T$ functions
\begin{align}
    \phi_t:\left(\Simplex\right)^{t-1}\to\Simplex,~~t=1,\ldots,T,
\end{align}
such that at the $t$th round, the house chooses $r_t=\phi_t(q^{t-1})$, where $q^{t-1}=(q_1,\ldots,q_{t-1})$ are the $t-1$ points in the simplex that the gambler chose in the previous rounds.
\item \textbf{Gambler's Strategy/Algorithm}: A strategy for the second player (the gambler) is a set of $T$ functions
\begin{align}
    \psi_t:\left(\Simplex\right)^{t}\to\Simplex,~~t=1,\ldots,T,
\end{align}
such that at the $t$th round, the gambler chooses $q_t=\psi_t(r^{t})$.
\item \textbf{Individual accumulated loss vector}: The individual accumulated  loss vector of a given house strategy $\{\phi_t\}=\{\phi_t\}_{t=1}^T$ and  $T$ points $q^T=(q_1,\ldots,q_T)$ chosen by the gambler is
\begin{align}
\varphi_T(\{\phi_t\},q^T)=\sum_{t=1}^T f(r_t=\phi_t(q^{t-1}),q_t). 
\end{align}
\item \textbf{Individual game loss}: The individual game loss of a given house strategy $\{\phi_t\}$ and the $T$ points $q^T$ chosen by the gambler corresponds to the largest return the gambler can make, and is defined as 
\begin{align}
\|\varphi_T(\{\phi_t\},q^T)\|_{\infty}=\max_{i=0,\ldots,m-1}\left\{\sum_{t=1}^T f_i(r_t=\phi_t(q^{t-1}),q_t)\right\}. 
\end{align}
\item \textbf{House strategy's loss}: The loss of a given house strategy $\{\phi_t\}$ is 
\begin{align}
L_T\left(\{\phi_t\}\right)=\max_{q^T\in(\Simplex)^T}\|\varphi_T(\{\phi_t\},q^T)\|_{\infty}.\label{eq:LTdef}
\end{align}
\item \textbf{Optimal house loss}: The optimal house loss is
\begin{align}\label{eq:opt_return}
L^*_T=\min_{\{\phi_t\}}L_T\left(\{\phi_t\}\right)= \min_{\{\phi_t\}}\max_{q^T\in(\Simplex)^T}\|\varphi_T(\{\phi_t\},q^T)\|_{\infty}.
\end{align}
\end{itemize}

When the underlying event has only $m=2$ possible outcomes, we refer to the game as the \emph{binary online bookmaking game}. This paper considers only the binary case. Note that for the case $m=2$, the simplex $\Simplex$ corresponds to the interval $[0,1]$, so the probability vectors $r_t,q_t\in\Simplex$ may be represented by a single number. With some abuse of notation, in the binary case we therefore use $r_t,q_t\in[0,1]$ to denote $r_t(1),q_t(1)$, respectively, and the optimal house loss is defined via
\begin{align}
    L_T^* = \min_{\{r_t\}} \max_{q^T\in[0,1]^T} \max\left\{\sum_{t=1}^T \frac{q_t}{r_t(q^{t-1}) }, \sum_{t=1}^T \frac{1-q_t}{1-r_t(q^{t-1})}\right\}
\end{align}
where in the binary case we denote the house strategy at time $t$, that assigns a number $r_t\in[0,1]$ to each vector $q^{t-1}\in[0,1]^T$, by $r_t(q^{t-1})$.

\medskip

We note that the online bookmaking game is related to the dynamic betting setup described above as follows: the largest gain the bookmaker can guarantee, regardless of the gambler's behavior and of the event's outcome, is $T(1-\frac{L_T^*}{T\Gamma})$. In particular, if $\Gamma>\frac{1}{T}L_T^*$, the bookmaker can guarantee a positive gain.

\medskip

\begin{remark}
Note that the definition of a gambler's strategy $\{\psi_t\}$ is not really needed for defining the house strategy's loss as well as the optimal house loss, as those quantities are computed by taking the worst-case sequence $q^T\in\Simplex^T$. We nevertheless chose to include the gambler's strategy in the online bookmaking game definition to facilitate the operational interpretation of this game as representing the online gambling procedure described above. 
\end{remark}

\medskip

The following theorem summarizes our main results, presenting the optimal house loss and the performance of two algorithms based on whether the gambler is decisive. Specifically, a \emph{decisive gambler} chooses $q^T \in \{0,1\}^T$, that is,  at each step it places its entire bet on a single outcome of the experiment.
\begin{theorem}\label{th:main}
The optimal house loss for the binary online bookmaking game is
\begin{align}\label{eq:opt_loss_theorem}
L^*_T=T+\sqrt{T}.    
\end{align}
Moreover, if the gambler is decisive, Algorithm \ref{alg:bets_for_bin_2} (Section \ref{sec:alg}) defines a house strategy $\{r_t^{\texttt{ALG}}\}=\{\phi_t^{\texttt{ALG}}\}$ that can be computed with $T$ simple operations, and achieves the optimal house loss
\begin{align}
    \| \varphi_T (\{r_t^{\texttt{ALG}}\},q^T) \|_\infty = L^*_T
\end{align}
for all $q^T\in\{0,1\}^T$. If the gambler is non-decisive, i.e., $q^T\in[0,1]^T$, the house strategy $\{\overline{r}_t^{\texttt{ALG}}\}=\{\overline{\phi}_t^{\texttt{ALG}}\}$ (defined in Eq.  \eqref{eq:strategy_continuous}) achieves an individual game loss that can be bounded as
$$\| \varphi_T(\{\overline{r}_t^{\texttt{ALG}}\},q^T)\|_\infty \le L^*_T$$
for all $q^T\in[0,1]^T$.
\end{theorem}
Recall that $L_T^*$ corresponds to the scenario where both the house and the gambler play optimally, and the event's outcome is chosen to maximize the gambler's gain. Even if the gambler plays optimally, the gambler's gain can be smaller than $L_T^*$ if the event's outcome does not play to its favor.

The main difference in the guarantees of Algorithm \ref{alg:bets_for_bin_2} and its expected version in \eqref{eq:strategy_continuous} for continuous bets stems from the gambler's behavior. We show in Section \ref{sec:proof} that decisive (binary) gamblers maximize the house's loss. Thus, when a bookmaker follows the optimal strategy in Algorithm \ref{alg:bets_for_bin_2} against an optimal gambler its return is exactly $L_T^*$. However, if the gamblers are not decisive, the house may benefit from such behavior so that its loss can be decreased. For instance, in the extreme case where the gambler distributes its money according to the offered bet, i.e., $q_t=r_t$, the house's loss can be can be as low as $T$.

\subsection{Related Work and Connections to Other Problems}
\label{subsec:relatedwork}

\textbf{Competing with fixed strategies:} Observe that the (non-causal) fixed strategy 
\begin{align}
r_t=\hat{q}=\frac{1}{T}\sum_{i=1}^T q_t,~~~t=1,\ldots,T    
\end{align} 
attains the individual accumulated loss vector $\varphi_T(\hat{q},q^T)=(T,\ldots,T)^\tr$. Thus, the objective of designing an algorithm $\{\phi_t\}$ that attains house strategy loss $L_T(\{\phi_t\})$ close to $T$, is identical to that of designing an algorithm that attains minimal regret with respect to the class of all fixed (time-invariant) strategies $r_t=r~:~\forall t=1,\ldots,T$, where $r$ runs through all points in the simplex $\Simplex$. One can also consider a discretization of $\Simplex$ with $M$ points $r^1,\ldots,r^M\in\Simplex$ and aim for a minimal regret with respect to only those $M$ ``experts''.  For scalar bounded loss functions , there are many online optimization algorithms that attain regret of $O(\sqrt{T\log M})$ with respect to any class of $M$ experts~\cite{cesa1997use}\cite{hazan2023introductiononlineconvexoptimization}, the multiplicative weights updates (MWU) algorithm being a canonical representative~\cite{arora2012multiplicative}. 
%\OrComm{I think that the MWU does not require convexity of the loss function or bounded gradients. It seems to only require that the loss is bounded. See Corollary 3 in Arora-Hazan-Kale.}
Note, however, that our online optimization problem is \emph{vector}-valued, and furthermore, the loss functions $f_0(r,q),\ldots,f_{m-1}(r,q)$ are unbounded in $\Simplex\times\Simplex$. The problem of unbounded loss may be handled by choosing $0<\delta_T<1$ and restricting the class of competing fixed strategies to the region $\Simplex\cap[\delta_T,1]^m$, such that for any $r$ in this region and any $q\in\Simplex$, we have that $\|f(r,q)\|_\infty\leq \frac{1}{\delta_T}$. The value of $\delta_T$ should be chosen small enough such that for any point $r\in\Simplex$ there is a ``close enough'' allowed point in $\Simplex\cap[\delta_T,1]^m$, but large enough so that the upper bound $\frac{1}{\delta_T}$ on $\|f(r,q)\|_{\infty}$ is not too large. When such a restriction is employed, our problem becomes a vector-valued online optimization problem with bounded loss. While methods such as MWU, designed for scalar online optimization, are not suitable for such a task, the Blackwell approachability framework may be applied.

\medskip

\textbf{Blackwell approachability:} For a vector-valued game with loss function $f(r,q)\in\RR^m$, Blackwell~\cite{blackwell1956analog} have posed and answered the following question (see~\cite[Definition 13.3]{hazan2023introductiononlineconvexoptimization}): \emph{Given a convex region $\m{S}\subset\RR^m$, can we find a strategy $\{\phi_t(q^{t-1})\}_{t=1}^T$ such that the Euclidean distance between $\frac{1}{T}\sum_{i=1}^T f(r_t=\phi_t(q^{t-1}),q_t)$ and $\m{S}$ vanishes with $T$, uniformly in $q^T$?} Here, a distance between a point $x\in\RR^m$ and a set $\m{S}\subset\RR^m$ is defined as $\min_{s\in\m{S}}\|x-s\|$. A set $\m{S}$ for which such a strategy exists is called \emph{approachable}, and Blackwell provided a simple necessary and sufficient condition for approachability, and provided an algorithm whose average loss vector $\frac{1}{T}\sum_{i=1}^T f(r_t=\phi_t(q^{t-1}),q_t)$ approaches $\m{S}$ uniformly for all $q^T$, provided that $\m{S}$ is approachable. Whenever the loss function is bounded, the convergence rate is $O(T^{-1/2})$.

In the online bookmaking game, we are interested in minimizing $\|\frac{1}{T}\sum_{i=1}^T f(r_t=\phi_t(q^{t-1}),q_t)\|_{\infty}$. Denoting the closed $\ell_{\infty}$ ball in $\RR^m$ by $B_{\infty}=\{x\in\RR^m~:~\|x\|_{\infty}\leq 1\}$, the question of characterizing the optimal house loss $L^*_T$ is equivalent to finding the smallest $\beta>0$ such that there exists a strategy $\{\phi_t(q^{t-1})\}_{t=1}^T$ for which $\frac{1}{T}\sum_{i=1}^T f(r_t=\phi_t(q^{t-1}),q_t)\in \beta\cdot  B_{\infty}$. Thus, in principle, one can use Blackwell's algorithm for designing a ``good'' house strategy. The problem, however, is that the loss function is not bounded for the online bookmaking game, and one must therefore restrict the strategy space to vectors in $\Simplex\cap[\delta_T,1]^m$ as described above. In Section~\ref{sec:Blackwell} we derive an algorithm which follows Blackwell's approach (taking into account the fact that $f(r,q)$ is unbounded). Such an algorithm only attains $\|\frac{1}{T}\sum_{i=1}^T f(r_t=\phi_t(q^{t-1}),q_t)\|_{\infty}=1+O(T^{-1/4})$ for all $q^T\in(\Simplex)^T$. This demonstrates that, to the best of our understanding, standard tools are insufficient for proving our main result, Theorem~\ref{th:main}, and do not even recover the optimal convergence rate. 

We note in passing that the objective in online bookmaking game is somewhat reminiscent to the objective in \emph{calibration}~\cite{foster1997calibrated, foster1998asymptotic, cesa2006prediction} which has been shown to be reducible to Blackwell approachability~\cite{foster1999proof, mannor2010geometric, abernethy2011blackwell}, as well as weak calibration~\cite{vovk2005defensive,kakade2004deterministic}. In particular, in the context of weak calibration, defining the two functions $w_1(r)=\frac{1}{r}$ and $w_2(r)=-\frac{1}{1-r}$, we see that a good strategy for the binary online bookmaking game guarantees that $\frac{1}{T}\sum_{t=1}^T w_i(r_t)(q_t-r_t)$ vanishes for both $i=1,2$. However, since these functions are not Lipschitz, an off-the-shelf strategy for weak calibration does not seem to imply a good strategy for the online bookmaking game, and vice versa. 

\medskip

\textbf{Dynamic programming:} The online bookmaking problem can be formulated as a Markov decision process (MDP) whose state is the vector that contains the accumulated bets on each team \cite{bertsekas}. The action at each time includes a min-max over $r_t,q_t$, respectively, while the action in the last time has an additional maximum over the accumulated bets' vector. It can be shown that this is a valid MDP so that actions that depend on the defined state are optimal. The defined MDP has the advantage that the actions and (normalized) states take values in  time-invariant spaces. However, the MDP suffers from the fact that those spaces are continuous and therefore even to evaluate the optimal gain numerically we need to use some approximation method, e.g., grid quantization over the states and actions. Interestingly, as we demonstrate, actions that depend on the entire history of actions (rather than just the current state) can simplify solutions to certain sequential problems.

\medskip

\textbf{Connection to universal compression:} The online bookmaking game is closely related to the universal compression problem. In particular, consider the universal compression problem~\cite[Section IV]{merhav1998universal} of designing a variable-length prefix-free source code $g:[m]^T\to\{0,1\}^*$ such that for any vector $x^T\in[m]^T$ we have that
\begin{align}
\ell(x^T)-T\cdot H(\hat{q}_{x^T})=o(T).    
\end{align}
Here, $\ell(x^T)$ is the length (number of bits) of the codeword $g(x^T)$ representing $x^T$, $\hat{q}_{x^T}$ is the empirical distribution (normalized histogram) of the sequence $x^T$, and $H(\cdot)$ is the entropy function (defined with  logarithm taken to base $2$, such that the entropy is measured in bits). It is well-known that the problem of designing a variable-length prefix-free code for sequences in $[m]^T$ is equivalent (up to constant number of bits) to setting a probability assignment on $[m]^T$~\cite{CoverThomas}, which in turn is equivalent to a sequence of $T$ conditional probability assignments $p_{X_t|X^{t-1}}:[m]^{t-1}\to\Simplex$, $t=1,\ldots,T$. In particular, we may use arithmetic coding with such conditional probability assignments and obtain~\cite{CoverThomas}
\begin{align}
\ell(x^t)\leq 2+\sum_{t=1}^T \sum_{i=0}^{m-1} \mathbbm{1}\{x_t=i\}\log_2\frac{1}{p_{X_t|X^{t-1}}(i|x^{t-1})}.
\end{align}
Any strategy $\{\phi_t\}$ for the online bookmaking game induces a sequence of $T$ conditional probability assignments $r_{X_t|X^{t-1}}:[m]^{t-1}\to\Simplex$, $t=1,\ldots,T$. To see this, assign to each value $i\in[m]$ the corresponding standard basis vector $\mathbf{e}_i$ (we index the coordinates from $0$ to $m-1$, and $\mathbf{e}_i$ is the vector whose $i$th coordinate equals $1$ and all the rest are zero). These vectors are points in $\Simplex$, so we may define (with some abuse of notation) the conditional probability assignments
\begin{align}
r_{X_t|X^{t-1}}(\cdot|x^{t-1})=\phi_t(q_1=\mathbf{e}_{x_1},\cdots,q_{t-1}=\mathbf{e}_{x_{t-1}}),~~t=1,\ldots,T.
\label{eq:univprobassignment}
\end{align}
In fact, we will see in the next section that the problem of designing an optimal strategy $\{\phi_t\}$ for the online bookmaking game is equivalent to that of designing an optimal strategy for the case where all points $q_1,\ldots,q_T$ are chosen from $\{\mathbf{e}_0,\ldots,\mathbf{e}_{m-1}\}$. With the probability assignment~\eqref{eq:univprobassignment} we have
\begin{align}
\sum_{t=1}^T \sum_{i=0}^{m-1} \mathbbm{1}\{x_t=i\}\log_2\frac{1}{r_{X_t|X^{t-1}}(i|x^{t-1})}&-T\cdot H(\hat{q}_{x^T})=\sum_{t=1}^T \sum_{i=0}^{m-1} \mathbbm{1}\{x_t=i\}\log_2\frac{\hat{q}_{x^T}(i)}{r_{X_t|X^{t-1}}(i|x^{t-1})}\label{eq:empent}\\
&=\sum_{t=1}^T \sum_{i=0}^{m-1} \mathbf{e}_{x_t}(i)\log_2\frac{\hat{q}_{x^T}(i)}{r_{X_t|X^{t-1}}(i|x^{t-1})}
\nonumber\\
&= \sum_{t=1}^T \log_2\left( \sum_{i=0}^{m-1} \mathbf{e}_{x_t}(i)\frac{\hat{q}_{x^T}(i)}{r_{X_t|X^{t-1}}(i|x^{t-1})} \right)\\
&\leq T \log_2\left( \sum_{i=0}^{m-1} \frac{1}{T}\sum_{t=1}^T\mathbf{e}_{x_t}(i)\frac{\hat{q}_{x^T}(i)}{r_{X_t|X^{t-1}}(i|x^{t-1})} \right)\label{eq:jens2}\\
&=T \log_2\left( \sum_{i=0}^{m-1}\hat{q}_{x^T}(i) \cdot \frac{1}{T}\sum_{t=1}^T\frac{\mathbf{e}_{x_t}(i)}{r_{X_t|X^{t-1}}(i|x^{t-1})} \right)\nonumber\\
&=T \log_2\left( \sum_{i=0}^{m-1}\hat{q}_{x^T}(i) \cdot \frac{1}{T}\sum_{t=1}^T f_i\left(r_t=\phi_t(\mathbf{e}_{x_1},\ldots,\mathbf{e}_{x_{t-1}}),\mathbf{e}_{x_t}\right)) \right)\nonumber\\
&\leq T \log_2\left(\frac{L_T(\{\phi_t\})}{T} \right),\label{eq:LTcompressionBound}
\end{align}
where~\eqref{eq:empent} follows since $\hat{q}_{x^T}(i)=\frac{1}{T}\sum_{t=1}^{T} \mathbbm{1}\{x_t=i\}$,~\eqref{eq:jens2} follows from Jensen's inequality and concavity of $y\mapsto \log_2(y)$, and~\eqref{eq:LTcompressionBound} follows from the definition of $L_T(\{\phi_t\})$ in~\eqref{eq:LTdef}. Thus, any strategy $\{\phi_t\}$ for the online bookmaking game with house strategy loss $L_T(\{\phi_t\})=T+o(T)$ can be translated to a universal compression scheme that attains redundancy of $o(T)$ bits with respect to the class of lossless compressors corresponding to an i.i.d. distribution. 

The opposite is not true. Assume the gambler is restricted to a \emph{decisive} strategy, where in each round it invests all its budget in a single outcome. This is equivalent to restricting the gambler to using $q^T\in\{\mathbf{e}_0,\ldots,\mathbf{e}_{m-1}\}^T$ rather than $q^T\in \Simplex^T$. A house strategy for such a gambler is a probability assignment $r_t(x^{t-1})=r_{X_t|X^{t-1}}(\cdot|x^{t-1})$ where $x_t$ is the outcome the gambler chose to bet on in round $t$. Probability assignments that give excellent redundancy for the universal compression problem may be very bad for the online bookmaking game. To see this, consider the binary case $m=2$ with the Shtarkov/Krichevsky-Trofimov probability assignment~\cite{krichevsky1981performance,shtar1987universal,merhav1998universal} $p_{X_t|X^{t-1}}(i|x^{t-1})=\frac{1/2+\sum_{j=1}^{t-1} \mathbbm{1}\{x_j=i\}}{t}$. While for universal compression this probability assignment yields redundancy of $O(\log T)$ bits for any $x^T\in\{0,1\}^T$~\cite[Chapter 13.5]{PW_book}, for a gambling sequence $x^T=0^{T-1} 1$, this probability assignment gives 
$$\sum_{t=1}^T f_1(r_t,q_t)=\frac{1}{p_{X_{T}|X^{T-1}}(1|0^{T-1})}=2T.$$
To summarize, any asymptotically good house strategy for the online bookmaking game provides an asymptotically good universal compression scheme, but asymptotically good universal compression schemes do not necessarily lead to asymptotically good house strategies for the online bookmaking game. Thus, in this sense, designing strategies for the online bookmaking game is a harder problem than designing schemes for universal compression.

\subsection{Paper Structure}
In Section~\ref{sec:alg} we present our optimal algorithm for decisive gamblers, and the algorithm it induces for general (non-decisive) gamblers. Section~\ref{sec:proof} provides the optimality proof of these algorithms, along with several observations on the structure of an optimal algorithm. While our optimal algorithm for decisive gamblers runs in linear time (in the horizon $T$), a brute-force implementation of the general algorithm it induces requires $O(2^T)$ operations. In Section~\ref{sec:montecarlo} we develop a polynomial-time algorithm that approximates the optimal one and attains loss arbitrarily close to that of the optimal one. In Section~\ref{sec:Blackwell} we develop a sub-optimal algorithm for the online bookmaking problem, based on the standard Blackwell-approachability approach. The loss of this algorithm is proved to be $T+O(T^{3/4})$, which is significantly larger than $L_T^*$. We nevertheless provide this algorithm for two reasons: 1)It demonstrates that standard approaches seem to be insufficient for attaining loss close to $L_T^*$; 2)While it is sub-optimal, it has one advantage over the algorithm from Section~\ref{sec:alg}: It does not depend on the horizon. We conclude and list several directions for future work in Section~\ref{sec:conc}.

\section{Algorithms for Optimal Bookmaking}\label{sec:alg}
In this section, we present two algorithms for the online bookmaking problem in which the experiment has two possible outcomes, i.e., the binary online bookmaking problem. We follow the terminology that the experiment's outcome is the winning team in a game between Team $0$ and Team $1$. The algorithms are sequential and take as input the sequence of bets $q_1q_2\dots q_{t-1}$ placed so far, and produce the sequence of odds $r_1r_2\dots r_t$ for $t=1,\dots,T$. Recall that the odds and the bet, at each time, are $r_t\in \Delta_m$ and $q_t\in \Delta_m$, and in the binary case each of them can be described with a scalar, i.e., $r_t\df r_t(1)\in[0,1]$ and $q_t\df q_t(1)\in[0,1]$. The corresponding pay-offs vector that will be published by the bookmaker at time $t$ is $[\gamma_t(0),\gamma_t(1)] = [\frac1{\Gamma (1-r_t)},\frac1{\Gamma r_t}]$.

The difference between the two algorithms is based on the gambler's behavior. First, we present the optimal algorithm for the case where the gambler is decisive, meaning their bets satisfy $q_t \in \{0,1\}$. For this case, Theorem \ref{th:main} asserts that, regardless of the gamblers' bets, $q^T$, the algorithm achieves the optimal loss $L_T^*$. We then extend the strategy in Algorithm \ref{alg:bets_for_bin_2} to handle continuous bets~$q_t \in [0,1]$.

\subsection{An Optimal Algorithm for Decisive Gamblers}
The algorithm for decisive bets is presented in Algorithm \ref{alg:bets_for_bin_2}. The main idea is to track two numbers, $(a, b)$, which serve as the state of the algorithm and are updated each time a new bet is placed. Initially, the state is set to the optimal loss, $(a,b) = (T + \sqrt{T}, T + \sqrt{T})$. 

The state variables $(a,b)$ correspond to the worst-case future losses, considering whether Team 0 or Team 1 will win the game, respectively. This point will be made clearer in the sequel. As new bets are placed, the state is updated using a specific formula that ensures that the worst-case losses for both teams are minimized. The derivation of the update formula and its optimality are provided in Theorem \ref{th:main_strategy}.

\begin{algorithm}
\caption{Optimal Strategy For Decisive Gamblers}
\label{alg:bets_for_bin_2}
\begin{algorithmic}
\State \textbf{Inputs:} $T$ (Total rounds), $q_1q_2\dots q_{T-1}$ (Bets are revealed sequentially)
\State \textbf{Output:} $r^T$ (House strategy)
\State \textbf{Initialization:} $a = T+ \sqrt{T}$, $b = T+ \sqrt{T}$
\State $r_1\gets \frac{1}{2}$
\For{$t=1:T-1$}
\State $d \gets T - t$
\If{$q_t = 0$} 
    \State $a \gets d \frac{b - (d-1)}{b - d}$ \Comment{Update the future loss for Team $0$}
\ElsIf{$q_t = 1$}
    \State $b \gets d \frac{a - (d-1)}{a - d}$ \Comment{Update the future loss for Team $1$}
\EndIf
\State $b^+ \gets (d-1) \frac{a - (d-2)}{a-(d-1)}$ \Comment{Hypothetical future loss for Team $1$}
\State $r_{t+1} \gets \frac1{b-b^+}$
\EndFor
\end{algorithmic}
\end{algorithm}
The strategy $r_{t+1}(q^t)$ only depends on $q^t$, and we denote the collection mappings produced by Algorithm \ref{alg:bets_for_bin_2} as $\{\phi^\texttt{ALG}_t\}=\{r^\texttt{ALG}_t\}$. The algorithm's implementation requires tracking two real numbers and at each cycle only a simple update is executed. 

\subsection{Algorithm for Non-Decisive Gamblers (Continuous Bets)}
\label{subsec:algcont}
In practice, the gambler is not a single entity. Rather, it represents the accumulation of bets made by several gamblers, that may distribute their bets over the two teams. We present a modification of Algorithm \ref{alg:bets_for_bin_2} to handle continuous bets $q_t\in[0,1]$. 
% \OrComm{This seems out of place here. I would delete it: ``Here, since the gamblers are not optimal in the sense of maximizing the bookmaker's loss, Theorem \ref{th:main} asserts that the loss is upper bounded by $T+\sqrt{T}$. In other words, the bookmaker can gain more, as illustrated in the example in the end of Section~\ref{subsec:mathform}.''}

The algorithm for continuous bets $q_t\in[0,1]$ is based on the optimal policy for binary bets, generated by Algorithm \ref{alg:bets_for_bin_2}. In particular, let $r_t^{\texttt{ALG}}(x^{t-1})$ denote the odds produced by Algorithm \ref{alg:bets_for_bin_2} for some input $x^{t-1}\in\{0,1\}^t$. The idea here is to view $q_t\in[0,1]$ as the expected value of a binary random variable $X_t\sim \text{Ber}(q_t)$. The strategy for continuous bets is   
\begin{align}\label{eq:strategy_continuous}
    \overline{r}^{\texttt{ALG}}_t(q^{t-1})&= \EE\left[r_t^{\texttt{ALG}}(X^{t-1})\right],
    % &\triangleq\textcolor{red}{\overline{\phi}^{\texttt{ALG}}_t(q^{t-1})},
\end{align}
where the expected value is taken with respect to the sequence of independent random variables $X_i\sim \text{Ber}(q_i)$. Note that \eqref{eq:strategy_continuous} defines a deterministic, instantaneous mapping, and we denote the sequence of mappings by $\{\overline{r}_t^{\texttt{ALG}}\}$.

Algorithm~\ref{alg:bets_for_bin_2} involves carrying out simple operations per round to compute the action $r_t^{\texttt{ALG}}(q^{t-1})$ when the gamblers are decisive, leading to an efficient total runtime of $\mathcal{O}(T)$. Unfortunately, this efficiency is lost when computing the odds for non-decisive gamblers. In this case, the odds at time $t$ are computed via~\eqref{eq:strategy_continuous} as  
\begin{align}
    \overline{r}^{\texttt{ALG}}_t(q^{t-1})
    &= \sum_{x^{t-1}\in\{0,1\}^{t-1}} r_t^{\texttt{ALG}}(x^{t-1}) \prod_{i=1}^{t-1} q_i^{x_i}(1-q_i)^{1-x_i}.
    \label{eq:strategy_continuous2}
\end{align}
Therefore, calculating the odds when the gamblers place continuous bets involves a summation over $2^{t-1}$ terms which is computationally very expensive. In Section~\ref{sec:montecarlo}, we provide a low-complexity algorithm to approximately calculate $\overline{r}^{\texttt{ALG}}_t(q^{t-1})$ and quantify this algorithm's approximation error. This algorithm approximates the expectation operation in~\eqref{eq:strategy_continuous} via a Monte Carlo simulation.

\section{Proof of Main Results}\label{sec:proof}

In this section, we prove our main results in Theorem \ref{th:main}: we derive the optimal house loss and show that Algorithm \ref{alg:bets_for_bin_2} is optimal for decisive gamblers, as well as the performance of its extension when generalizing to continuous gamblers. Our technical results are more general than required for the online bookmaking setup, so we will present them in their full generality and prove Theorem \ref{th:main} at the end of this section.

\subsection{Decisive Gamblers are Optimal}\label{subsec:proof_bin_opt}
We first show that the house strategy's loss is achieved with gamblers that are decisive, that is, choosing a single team to bet on at each round is optimal from the gambler's perspective. In other words, if the gambler aims to maximize the house's loss it does not need to split its money among the teams and can place its bet on a single team in each round. It turns out that this claim holds not only for the online bookmaking problem, but for more general loss functions, defined as follows.
\begin{problem}[Convex-Linear-Max Games]\label{problem}
We consider the repeated game with two players as defined in Section \ref{subsec:mathform}, but allow more general loss functions than the ones in \eqref{eq:def_fi_intro}. Specifically, each loss function is taken from a finite collection $f_i(r,q) = q(i)g_i(r)$, where $g_i:\mathbb{R}^m\to\mathbb{R}$ is a convex function and $i\in[m]$. The optimal loss is defined as before,
\begin{align}\label{eq:prob_def_obj}
   L^*_T&= \min_{\{\phi_t\}} \max_{q^T\in\Simplex^T} \|\varphi_T(\{\phi_t\},q^T)\|_\infty.
\end{align}
\end{problem}
The online bookmaking problem is a special case of Problem \ref{problem} by choosing $g_i(r) = \frac1{r(i)}$. The optimal loss in \eqref{eq:prob_def_obj} is difficult to compute directly for several reasons such as the continuous domain of the players' strategies, the dependence of the mappings' domain on time, the inner maximization (the $\ell^\infty$ norm) etc. Our first question is whether the strategies' structure can be simplified for Problem~\ref{problem}.

One suggestion is to formulate the repeated game as a Markov decision process (MDP) whose state at time $t$ is the vector of cumulative past losses $s_t(i) = \sum_{j=1}^{t-1} f_i(r_j,q_j) $ for $i=1,\dots,m$. The MDP formulation benefits the fact that the actions $r_t$ and $q_t$ should not depend on the entire history of actions, $r^{t-1},q^{t-1}$, whose lengths grow with time. Instead, they only depend on the vector $s_t$. However, the challenge in solving this MDP is that the state is continuous, and that the overall loss function cannot be expressed as a sum of immediate rewards due to the infinity norm. Intuitively, the inner maximization in \eqref{eq:prob_def_obj} can be interpreted as being performed at the end of the game, as it depends on the vectors $r^T$ and $q^T$. Consequently, the players do not know during the game which specific function will ultimately be selected at the end.

We take a different approach in which we aim to restrict the strategies' domains and images. We prove that the image of the second player's strategy, defined by $\psi_t$, can be restricted to the simplex vertices, i.e., the basis vectors $\{\mathbf{e}_i\}_{i=1}^m$. Formally, define the set of restricted strategies as
\begin{align}\label{eq:psi_limited}
    \psi^{\mathtt{D}}_t: (\Simplex)^t \to \{\mathbf{e}_1,\dots\mathbf{e}_m\}
\end{align}
such that $q_t = \psi^{\mathtt{D}}_t(r^t)$. We call a player whose image is restricted as \eqref{eq:psi_limited} a \emph{decisive player} as it allocates all its mass to a single coordinate of the vector $q_t$. If the second player is decisive, it follows directly that the domain of the first player's strategy can be restricted as
\begin{align}\label{eq:phi_limited}
    \phi^{\mathtt{D}}_t: \{\mathbf{e}_1,\dots\mathbf{e}_m\}^{t-1}\to \Simplex
\end{align}
with $r_t = \phi^{\mathtt{D}}_t(q^{t-1})$ since values outside of this domain will not appear in $q^{t-1}$. We call a strategy of the form \eqref{eq:phi_limited} a \emph{skeleton strategy}. 

As mentioned, it will be shown that decisive players are optimal so that skeleton strategies suffice to achieve the optimal loss. However, one may be interested in algorithms for non-decisive players, i.e., $q^T\in\Simplex^T$. For this purpose, we can take any skeleton strategy, and define its expected skeleton strategy as
\begin{align}\label{eq:ex0_skeleton}
    \overline{\phi}_t(q^{t-1})&= \E[\phi^{\mathtt{D}}_t(X^{t-1})],
\end{align}
where $X^{t-1}$ is a sequence of independent random variables with $\Pr(X_t = \mathbf{e}_i) = q_t(i)$. It will be shown that, if the second player is non-decisive, the first player can only gain from such behavior by utilizing the strategy in \eqref{eq:ex0_skeleton}. The results of this section are formalized as follows. 

\begin{theorem}\label{th:discrete_is_opt}
The optimal loss of the convex-linear-max game in Problem \ref{problem} can be achieved with a decisive player and skeleton strategies. That is, the optimal loss can be expressed as
\begin{align}\label{eq:th_decisive}
   L^*_T&= \min_{\{\phi^\mathtt{D}_t\}} \max_{q^T\in\{\mathbf{e}_1,\dots,\mathbf{e}_m\}^T} \| \varphi_T(\{\phi^\mathtt{D}_t\},q^T)\|_\infty.
\end{align}
Moreover, any skeleton strategy, $\{\phi^{\mathtt{D}}_t\}$, and its expected skeleton strategy, $\{\overline{\phi}_t\}$, satisfy
\begin{align}\label{eq:th_decisive_ub_expected}
\|\varphi_T(\{\overline{\phi}_t\},q^T)\|_\infty&\le \max_{x^T\in\{\mathbf{e}_1,\dots,\mathbf{e}_m\}^T}\|\varphi_T(\{\phi^{\mathtt{D}}_t\},x^T)\|_\infty
\end{align}
for all $q^T\in\Simplex^T$.
\end{theorem}

\begin{proof}[Proof of Theorem \ref{th:discrete_is_opt}]
We prove that the loss incurred by skeleton strategies and decisive players, i.e., the right-hand side (rhs) of \eqref{eq:th_decisive}, serves as both an upper and a lower bound for the optimal loss $L_T^*$. A byproduct of the upper bound derivation is the inequality in \eqref{eq:th_decisive_ub_expected}, which provides a performance bound on the expected skeleton strategy.

Restricting the maximization in \eqref{eq:prob_def_obj} to $q^T \in \{\mathbf{e}_1, \dots, \mathbf{e}_m\}^T$ implies that the minimization over strategies can be restricted to skeleton strategies $\{\phi^{\mathtt{D}}_t\}$, which leads to the lower bound:
\begin{align}
    \min_{\{\phi^{\mathtt{D}}_t\}} \max_{q^T \in \{\mathbf{e}_1, \dots, \mathbf{e}_m\}^T} \| \varphi_T(\{\phi^{\mathtt{D}}_t\}, q^T)\|_\infty &\leq L_T^*.
\end{align}

For the upper bound, we have for any strategy $\{\phi_t\}$, 
\begin{align}\label{eq:proof_deci_UB_any}
    L^*_T &\le \max_{q^T\in\Simplex^T} \| \varphi_T(\{\phi_t\},q^T)\|_\infty.
\end{align}
We choose the strategy in \eqref{eq:proof_deci_UB_any} as the expected skeleton strategy in \eqref{eq:ex0_skeleton} of some skeleton strategy $\{\phi^\mathtt{D}_t\}$, and show that the maximum over $q^T\in\Simplex^T$ is attained at $q^T\in \{\mathbf{e}_1,\dots\mathbf{e}_m\}^T$.

Consider the following upper bound on the rhs of \eqref{eq:proof_deci_UB_any} as a function of $q^T\in\Simplex^T$:
\begin{align}\label{eq:proof_discrete_main}
    \| \varphi_T(\{\overline{\phi}_t\},q^T)\|_\infty &\stackrel{(a)}= \max_{i\in[m]} \sum_{t=1}^T q_t(i)g_i(\E[\phi^\mathtt{D}_t (X^{t-1})]) \nonumber\\
    &\stackrel{(b)}\le \max_{i\in[m]} \sum_{t=1}^T q_t(i) \E[g_i(\phi^\mathtt{D}_t(X^{t-1}))]) \nonumber\\
    &= \max_{i\in[m]} \sum_{t=1}^T \E[\mathbbm{1}\{X_t= \mathbf{e}_i\}] \E[g_i(\phi^\mathtt{D}_t(X^{t-1}))]) \nonumber\\
    &\stackrel{(c)}= \max_{i\in[m]} \E\left[\sum_{t=1}^T \mathbbm{1}\{X_t = \mathbf{e}_i\} g_i(\phi^\mathtt{D}_t(X^{t-1}))\right]\nonumber\\
    % &\le \textcolor{red}{\max_{\tilde{q}^T\in\Simplex^T}}\E\left[\max_{i\in[m]} \sum_{t=1}^T \mathbbm{1}\{X_t = \mathbf{e}_i\} g_i(\phi^\mathtt{D}_t(X^{t-1}))\right]\nonumber\\
    &\stackrel{(d)}\le \max_{x^T\in\{\mathbf{e}_1,\dots\mathbf{e}_m\}^T}\max_{i\in[m]} \sum_{t=1}^T \mathbbm{1}\{x_t = \mathbf{e}_i\} g_i(\phi^\mathtt{D}_t(x^{t-1}))\nonumber\\
    &= \max_{x^T\in\{\mathbf{e}_1,\dots\mathbf{e}_m\}^T} \|\varphi_T( \{\phi^\mathtt{D}_t\}, x^T)\|_\infty,
\end{align}
where $(a)$ follows from the definition of the expected skeleton strategy in \eqref{eq:ex0_skeleton}, $(b)$ follows from Jensen's inequality, and $(c)$ follows from the independence of $X_t$ and $X^{t-1}$. Step $(d)$ follows from the fact that ${q}^T$ affects the expected value only via the probabilities allocated to each realization $x^T\in\{\mathbf{e}_1,\dots,\mathbf{e}_m\}^T$. Thus, to maximize the expected value, $q^T$ can be chosen such that $\Pr(X^T=x^T) = 1$ for any tuple $x^T$ that maximizes the argument of the expectation. The proof of the upper bound is completed by recalling that \eqref{eq:proof_discrete_main} holds for any skeleton strategy $\{\phi^\mathtt{D}_t\}$ so we can take a minimum over the latter. 
\end{proof}

\subsection{Optimal Solution via Balancing Trees} 
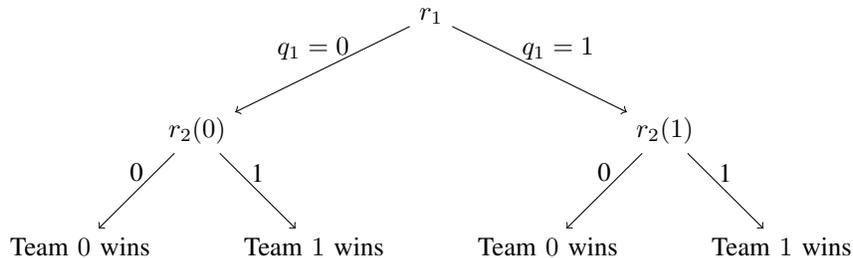
\begin{figure}[b!]
    \centering
\begin{forest}
for tree={
    grow=south, % Tree grows from left to right
    edge={->}, % Arrowed edges
    s sep=10mm, % Sibling separation
    l sep=10mm, % Level separation
    edge label={node[midway,above,font=\scriptsize]{#1}}, % Labels on edges
}
[$r_1$
    [$r_2(0)$, edge label={node[midway,above]{$q_1=0\ \ $}}
        [Team $0$ wins, edge label={node[midway,above]{0}}]
        [Team $1$ wins, edge label={node[midway,above]{1}}]
    ]
    [$r_2(1)$, edge label={node[midway,above]{$\ \ \ \   q_1=1$}}
        [Team $0$ wins, edge label={node[midway,above]{0}}]
        [Team $1$ wins, edge label={node[midway,above]{1}}]
    ]
]
\end{forest} 
\caption{A binary tree that describes the online bookmaking setup with $T=2$. First, the house chooses $r_1$, followed by the gambler's bet $q_1\in\{0,1\}$. The house then chooses the  odd $r_2(q_1)$ based on the bet $q_1$, and lastly, the bet $q_2$ is placed. As shown in Claim \ref{claim:winning_horse}, the last bet $q_2$ is equal to the winning team.}
    \label{fig:binarytree}
\end{figure}

In the previous section, we showed that the optimal house loss can be achieved with house skeleton strategies and decisive gamblers. In the case of binary games, decisive gamblers simply choose $q^T\in \{0,1\}^T$, and the optimal house loss in \eqref{eq:opt_return} is simplified to 
\begin{align}\label{eq:balancing_main}
    L_T^*&= \min_{\{r_t\}} \max_{q^T\in \{0,1\}^T} \max \left\{\sum_{t=1}^T \frac{1-q_t}{1-r_t(q^{t-1})}, \sum_{t=1}^T \frac{q_t}{r_t(q^{t-1})}\right\},
\end{align}
where $r_t:\{0,1\}^{t-1}\to[0,1]$, $t=1,\ldots,T$ denotes the the house strategy in the binary case. It is convenient to represent the interaction between the house and the gambler as a complete binary tree of depth $T$, where each node is associated with an odd $r_t \in [0,1]$, and edges correspond to the gambler's bets, $q_t \in \{0,1\}$. We illustrate this interaction in Fig. \ref{fig:binarytree} for $T = 2$. The difficulty in solving this optimization problem lies in the fact that the loss function remains unknown to the house and the gambler during the interaction. This is reflected in the inner maximization in \eqref{eq:balancing_main}, which is taken at the end of the interaction. Our solution is based on several simplifying claims that lead to the development of our new methodology, referred to as bi-balancing trees. The first claim concerns the gambler's bet in the final round.

\medskip

\begin{claim}\label{claim:winning_horse}
An optimal gambler places its final bet, $q_T$, on the winning team. As a result, the optimal house loss can be written as
\begin{align}\label{eq:optloss_claim1}
 L_T^*&= \min_{\{r_t\}} \max_{q^T\in \{0,1\}^T} g(\{r_t\},q^T),
 \end{align}
 where
 \begin{align*}
 g(\{r_t\},q^T)=\mathbbm{1}{\{q_T=0\}}\sum_{t=1}^T \frac{1-q_t}{1-r_t(q^{t-1})}+\mathbbm{1}{\{q_T=1\}}\sum_{t=1}^T \frac{q_t}{r_t(q^{t-1})}.    
\end{align*}
\end{claim}

\medskip

This claim may seem counterintuitive, as the final bet is placed before the outcome is known, however, both the gambler and the choice of the winning team act as adversaries and can therefore be viewed as a single player.

\begin{proof}[Proof of Claim \ref{claim:winning_horse}]
For a fixed strategy and bets sequence $q^{T-1}$, the maximization over $q_T$ and the inner maximization over the sums in \eqref{eq:balancing_main} can be exchanged. As $q_T=0$ and $q_T=1$ maximize the sums $\sum_{t=1}^T \frac{1-q_t}{1-r_t(q^{t-1})}$ and $\sum_{t=1}^T \frac{q_t}{r_t(q^{t-1})}$, respectively, we can write the maximum between the two optimized sums as $\max_{q_T}g(\{r_t\},q^T)$.
\end{proof}

We illustrate the structure of the online bookmaking problem and highlight the  challenges by solving the special case $T=2$.
\begin{example}
The optimization for $T=2$ is illustrated in Fig. \ref{fig:binarytree}. The strategy $\{r_t\}$ consists of three odds: $r_1$ at the root, and $r_2(0), r_2(1)$ that are chosen based on the branches labeled with $q_1\in\{0,1\}$. The challenge is that the house should choose the odds without the knowledge of the loss function. For example, after observing $q_1=0$, the house chooses $r_2(0)$, and the incurred loss can be either $\frac1{r_2(0)}$ or $\frac1{1-r_2(0)}$.

We write the possible losses for the optimization in \eqref{eq:optloss_claim1} as follows:
\begin{align}\label{eq:example}
  g(\{r_t\},q^2=00)&= \frac1{1 - r_1}   + \frac1{1 - r_2(0)}\nonumber\\
  g(\{r_t\},q^2=01)&= \frac1{r_2(0)} \nonumber\\
  g(\{r_t\},q^2=10)&= \frac1{1 - r_2(1)}\nonumber\\
  g(\{r_t\},q^2=11)&= \frac1{r_1} + \frac1{r_2(1)}.
\end{align}

As the gambler is taking a maximum over $q^2$, the best house strategy should be a one that equates the possible losses in \eqref{eq:example} (if it is possible). To see this, assume that $r_1,r_2(q_1)$ is an optimal strategy such that the losses are not equal. We have the chain $01\stackrel{r_2(0)}\iff 00\stackrel{r_1}\iff 11\stackrel{r_2(1)}\iff 10$ in which each link corresponds to a single odd that only affects the neighboring losses in a different direction. For instance, the link labeled with $r_2(0)$ implies that $g(\{r_t\},q^2=01)$ is a decreasing function, while $g(\{r_t\},q^2=00)$ is an increasing function of $r_2(0)$. If the strategy is not equalizing there is a chain of links that connects a bets sequence with the maximal loss to a sequence that has a smaller loss. We can then gradually decrease the maximum by perturbing each odd on the links starting from the lower loss towards the maximal loss. 

Direct calculation shows that $r_1 = \frac1{2}, r_2(0) = 1 - \frac1{\sqrt{2}}, r_2(1) = \frac1{\sqrt{2}}$ is the only strategy that equates the four losses to $2+\sqrt{2}$. This provides us with the optimal strategy and the optimal house loss $L_2^* = 2 + \sqrt{2}$. Indeed, the derivation shows that, if the winning team is equal to the final bet of the gambler, the loss is equal to the optimal house loss no matter what is the bets sequence.
\end{example}
From the above example, we deduce another observation. 

\medskip

\begin{claim}\label{claim:equalizing}
    If there exists a strategy that equalize the losses $g(\{r_t\},q^T)$ (for all $q^T \in \{0,1\}^T$), then an optimal strategy is among those that equalize the losses.
\end{claim}

\medskip

The proof of Claim \ref{claim:equalizing} is deferred to the end of the section, as it requires several definitions that will only be given later. 

The objective now is to find a strategy $\{r^T\}$ that equates the losses in \eqref{eq:optloss_claim1} for all bets sequences $q^T$ (if such a strategy exists). The direct approach to find a strategy that equates all bets' sequences is too involved even for small horizons, e.g., $T=3,4$. Our main idea is to study a more general problem which we call the bi-balancing problem: find a strategy that simultaneously equates the accumulated losses for each team. By Claim \ref{claim:winning_horse}, it reduces to equating the accumulated loss $\sum_{t=1}^T \frac{1-q_t}{1-r_t(q^{t-1})}$ for sequences with $q_T=0$, and the accumulated loss $\sum_{t=1}^T \frac{q_t}{r_t(q^{t-1})}$ for sequences with $q_T=1$.

To formalize the bi-balancing problem, we define the individual value function.

\medskip

\begin{definition}[Individual value function]
For a fixed strategy $\{r_t\}$, the \emph{individual value function} is the vector $(V^0(q^t),V^1(q^t))$ consisting of 
\begin{align}\label{eq:def_indvalfunction}
   V^0(q^t)&= \sum_{i=1}^t \frac{1-q_i}{1-r_i(q^{i-1})} \nonumber\\
   V^1(q^t)&= \sum_{i=1}^t \frac{q_i}{r_i(q^{i-1})}. 
\end{align}
\end{definition}
% \OrComm{Currently this definition applies only to the root of the binary tree. I think you want to introduce already here the notation $V^i_{q^j}(q_{j+1}^{T})$ that you introduce below.} \os{In this sub-section we do not use the prefix in the subscript. We only use in the optimal strategy derivation and Claim 2 proof in the next two sections. If you think it is clearer here, we can say that if subscript is omitted it corresponds to a null sequence which means root of a tree. }
\medskip

In words, the individual value function vector consists of the two possible future losses assuming that the forthcoming bets sequence is $q^t$. Note that the individual value function is defined for a tree with an arbitrary depth $t$ that is not necessarily equal to the horizon $T$. By Claim \ref{claim:winning_horse}, $g(\{r_t\},q^T)=V^i(q^T)$ for all sequences that end with $q_T=i$ for $i=0,1$. Thus, we  partition the sets of binary sequences as $\mathcal Q^T_i = \{ q^T\in \{0,1\}^T: q_T=i\}$. We can now define the bi-balancing problem.

\medskip

\begin{definition}[Bi-Balanced Trees]\label{def:bi-balan}
% A strategy is said to \emph{bi-balance} a complete binary tree if the individual value function $(V^0(q^T),V^1(p^T))$ is constant for all $q^T\in \mathcal Q^T_0$ and $p^T\in \mathcal Q^T_1$. The collection of all vectors $(V^0,V^1)$ such that $V^i = V^i(q^T)$ is constant called the set of \emph{achievable value functions}. 
A complete binary tree is \emph{bi-balanced} by the strategy $\{r_t\}$ if the individual value function $(V^0(q^T),V^1(p^T))$ is constant for all $q^T\in \mathcal Q^T_0$ and $p^T\in \mathcal Q^T_1$. The collection of vectors $(V^0,V^1)$ such that $V^i = V^i(q^T)$ remains constant, for some strategy $\{r_t\}$ and all $q^T\in\mathcal Q^T_i$, is referred to as the set of \emph{achievable value functions}.
\end{definition}

\medskip

A solution to the bi-balancing problem specializes to a solution for the online bookmaking problem if we also require $V^0(q^T) = V^1(p^T)$ for some $q^T\in \mathcal Q^T_0$ and $p^T\in \mathcal Q^T_1$, that is, an identical house loss for all bets sequences. Also, Definition \ref{def:bi-balan} focuses on complete binary trees, while the problem formulation can be trivially generalized to non-binary trees, and even to non-complete trees if some bets sequences are assumed to be not feasible.

Our approach to the bi-balancing problem is to parameterize the set of achievable value functions on any node of the binary tree. Specifically, an explicit formula that relates the vector coordinates of the individual value function will be given using the function 
\begin{align}\label{eq:fd}
 f_{d}(x) = d \frac{x-(d-1)}{x-d},
\end{align}
where $d$ is the tree depth.
Note that $f_d(x)$ is an involution, i.e., $f_d(f_d(x))=x$, a key property that will be used in the sequel.
\begin{figure}[b]
    \centering
    \includegraphics[width=0.4\textwidth]{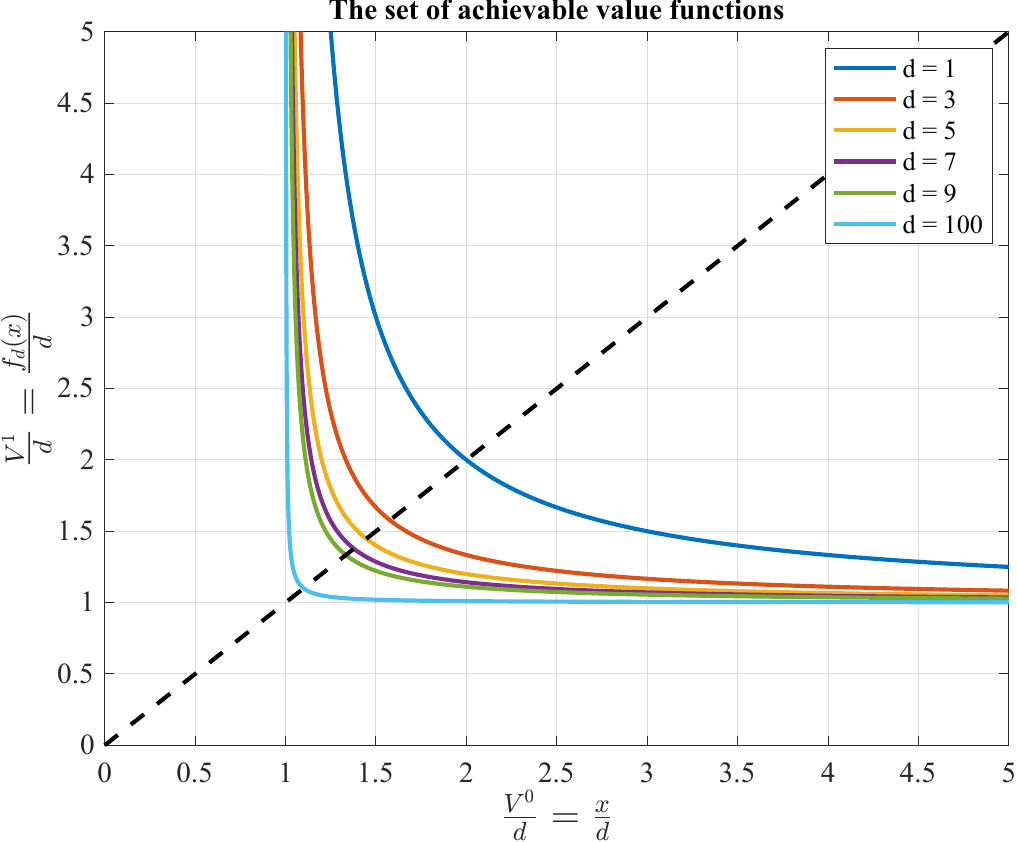} % Replace with your file name and desired width
    \caption{The set of achievable value functions in Theorem \ref{th:bi-balanced} as a function of the remaining depth, normalized by depth. The curves are symmetric around the line $V^1=V^0$ due to symmetry between the teams. The optimal house loss for an horizon $d$ can be obtained by intersecting its curve with the dashed line $V^0 = V^1$. As $d$ grows, the intersection approaches the lower bound $(V^0,V^1) = (1,1)$ with a convergence rate of $\frac1{\sqrt{d}}$.}
    \label{fig:your_label}
\end{figure}

We first characterize the set of achievable value functions and then focus on deriving the optimal strategy in Section \ref{subsec:opt_stra}.
The following is our main result regarding the bi-balancing problem. 

% It should be clear that one can compute the individual value function for each node \OrComm{it's not clear what the individual value function of a \emph{node} means} on a binary tree using a house strategy and vice versa. 

\medskip
 
\begin{theorem}\label{th:bi-balanced}
    For a tree of depth $d$, the set of achievable value functions is characterized by the union of vectors $(f_d(x),x)$ over $x > d$. That is, the individual value function of any bi-balanced tree of depth $d$ has the form $(f_d(x),x)$ for some $x>d$. Moreover, for a fixed $x>d$, there exists a single strategy whose individual value function is $(f_d(x),x)$.
\end{theorem}

\medskip

We use Theorem \ref{th:bi-balanced} to build a bi-balanced tree in a bottom-up fashion, by connecting two bi-balanced trees of depth $d$ to a single root, ensuring that the new tree, whose depth is $d+1$ is also bi-balanced. Theorem \ref{th:bi-balanced} provides a complete characterization for the relation between the coordinates of the individual value function; for any horizon $T$ where $d$ steps remain, if the future loss for the first team is $x$, then $f_d(x)$ is the future loss for the second team. 
% \OrComm{I think you are essentially describing eq.~\eqref{eq:proof_bibala2} and~\eqref{eq:proof_bibalan} here, but it's difficult to understand the intention without seeing those equations first. Should we tell the reader already here that the individual value of any path $q^T$ can be broken to 
% \begin{align*}
% V^0(q^T)&=V^0(q^d)+\mathbbm{1}\{q_{d+1}=0\}\left[\frac{1}{1-r_{d+1}(q^d)}+V_{q^d 0}^i(q_{d+2}^T) \right]+\mathbbm{1}\{q_{d+1}=1\}V_{q^d 1}^i(q_{d+2}^T) \\
% V^1(q^T)&=V^1(q^d)+\mathbbm{1}\{q_{d+1}=0\}V_{q^d 0}^i(q_{d+2}^T)+\mathbbm{1}\{q_{d+1}=1\}\left[\frac{1}{r_{d+1}(q^d)}+V_{q^d 1}^i(q_{d+2}^T) \right]
% \end{align*}}
The claim is symmetric if we exchange the roles of the first and the second teams; in particular, the set of achievable functions can also be characterized by the union of $(x,f_d(x))$. This can be easily deduced from the fact that $f_d(x)$ is an involution.

The value function of a bi-balanced tree depends on two parameters, the remaining tree depth $d$ and the argument $x$. It does not depend directly on $T$, but only via $x$. Particularly, for a binary tree of height $T$, there are $2^{T-d}$ nodes whose remaining depth is $d$. Clearly, each node at this  depth has a different value function, and Theorem \ref{th:bi-balanced} does not provide the individual value functions for each tree node (this will be provided later). Still, it is interesting to note that the value function structure is identical, at any fixed depth, no matter what is the overall horizon $T$ and what is the prefix sequence leading to the sub-tree. The set of achievable value functions is illustrated in Fig.~\ref{fig:your_label} where, for improved visibility, the normalized value function $\frac{1}{d}(f_d(x), x)$ is plotted.

As mentioned above, our derivation relies on a recursive construction of a bi-balanced tree from the bottom to the top. Specifically, we employ induction on the tree depth, where at each step two bi-balanced trees are connected by linking their roots to a new common root. This  provides a subset of the achievable value functions, as it is based on a particular construction.

Before proceeding to the proof of the theorem, we claim the converse: our construction captures any bi-balanced tree. The left and right sub-trees of a tree with depth $d$ are defined as the two trees of depth $d-1$ connected to the tree's root with $q_1 = 0$ and $q_1 = 1$, respectively. For a strategy $\{r_t\}_{t=1}^d$, the individual value function of the left sub-tree is a vector that consists  
\begin{align}
    V^0_L(q_2^d)&= \sum_{t=2}^d \frac{1 - q_t}{1 - r_t(0q_2^{t-1})}\nonumber\\
    V^1_L(q_2^d)&= \sum_{t=2}^d \frac{q_t}{r_t(0q_2^{t-1})},
\end{align}
and similarly for the right sub-tree, with $q_1 = 0$ replaced by $q_1 = 1$. The optimality of our construction follows from the following claim. 
\begin{claim}\label{claim:subtrees}
If a tree of depth $d$ is bi-balanced, then its two descendant sub-trees are bi-balanced. Specifically, the values $V_L^0(q_2^d)$, $V_R^0(q_2^d)$, $V_L^1(p_2^d)$, and $V_R^1(p_2^d)$ remain constant for all $q_2^d \in \mathcal{Q}_0^{d-1}$ and $p_2^d \in \mathcal{Q}_1^{d-1}$.
\end{claim}
This is analogous to the dynamic programming principle, where a problem can be solved optimally by optimally solving its tail subproblems. 

\begin{proof}[Proof of Claim \ref{claim:subtrees}]
If any of the sub-trees is not bi-balanced, there exist two sequences $q_2^{d}, \tilde{q}_2^{d} \in \mathcal{Q}^{d-1}_0$ such that $V_L^0(q_2^d) \neq V_L^0(\tilde{q}_2^d)$ (the choice of $V_L^0(\cdot)$ is without loss of generality). Then, it follows by \eqref{eq:def_indvalfunction} that $V^0(0 q_2^d) \neq V^0(0 \tilde{q}_2^d)$.
\end{proof}

\begin{proof}[Proof of Theorem \ref{th:bi-balanced}]
We construct a bi-balanced tree whose value function is $(x,f_d(x))$, and then claim the converse: any bi-balanced tree has the particular value function structure constructed by our method.

For the construction, we propose a sequential method that is inspired by Markov decision processes. The main idea is to construct a bi-balanced tree by properly connecting two bi-balanced sub-trees. This allows us to construct a bi-balanced tree using an induction on the depth $d=1,2,\dots$. The induction proof:
\begin{enumerate}
    \item \textbf{Base Case:} Consider a tree of depth $d=1$. The house strategy consists of a single odd $r_1\in[0,1]$ at the root, and is followed by two possible bets $q_1\in\{0,1\}$, each corresponds to a branch. If the individual value function on the left branch is $V^0(0) = x$, the odd satisfies $\rfrac{r_1} = x$, which implies in turn that $V^1(1) = \frac1{r_1} = f_1(x)$ as required. The constraint $r_1\in(0,1)$ implies $x > 1$.
    \item \textbf{Induction Hypothesis:} By the induction hypothesis, for any $x>d$, there exists a unique bi-balanced tree of depth $d$ whose value function is $(x,f_{d}(x))$.
    \item \textbf{Induction Step:} From the induction hypothesis, there exist two bi-balanced trees of depth $d$ with value functions $(f_d(y), y)$ and $(x, f_d(x)) $, for some $x, y \geq d$.\footnote{Without loss of generality, a value functions $(y,f_d(y))$ can be written as $(f_d(y),y)$ since the mapping $f_d(\cdot)$ is an involution, i.e., $f_d(f_d(y))=y$.} The trees with the value function $(f_d(y), y)$ and $(x,f_d(x))$ will be used as the left and right sub-trees, respectively. That is, a new root will be connected via the branch $q_1=0$ to the left sub-tree, and via the branch $q_1=1$ to the right sub-tree. At the root of the constructed tree, there is a new odd $r_1$, to be chosen such that the constructed tree is bi-balanced. The tree construction is illustrated in Fig. \ref{fig:construction}.
    
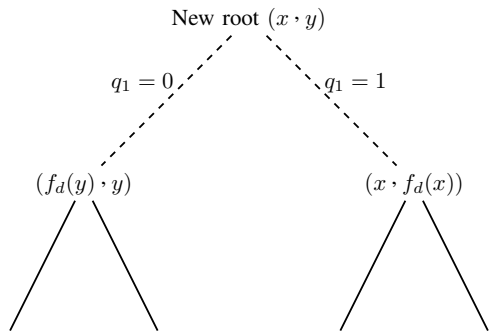
\begin{figure}[h!]
    \centering
    \resizebox{0.4\textwidth}{!}{ % This will scale the entire figure including spacing but keep the font size unchanged
    \begin{forest}
        for tree={
            grow=south,
            draw=none, % No drawing of borders around nodes
            minimum size=1.5em, inner sep=2pt,
            l sep=2cm, s sep=2cm,
            edge={black, thick},
            align=center,
            text height=1.5ex, text depth=.25ex
        }
        [New root $(x\CommaPunct y)$ 
            [$(f_d(y)\CommaPunct y)$, edge label={node[midway,above]{$q_1=0\ \ \ \ \ \ $}}, edge={dashed}
                [ ]
                [ ]
            ]
            [$(x\CommaPunct f_d(x))$, edge={dashed}, edge label={node[midway,above]{$\ \ \ \ \ \ q_1=1$}}
                [ ]
                [ ]
            ]
        ]
    \end{forest}
    }
    \caption{Illustration of the construction: a new root is connected with two dashed edges to two sub-trees, where each is bi-balanced with value functions $(f_d(y),y)$ and $(x,f_d(x))$. The sub-trees have a depth $d$ and the constructed tree is of depth $d+1$. The value function at the new root, $(x,y)$, follows from the fact that $q_1=0/1$ do not change the loss if team $1/0$ wins. The proof shows that there exists a unique odd at the root, $r_1$, such that the tree is bi-balanced and $y = f_{d+1}(x)$.}
    \label{fig:construction}
\end{figure}

    %We proceed to compute the individual value function at the root by utilizing their recursive relation with the individual value functions at the roots of the sub-trees.
    We proceed to compute the individual value function for the new $d+1$-deep tree by utilizing their recursive relation with the individual value functions of the $d$-deep sub-trees.
    We first consider sequences of the form $ 0q^{d}$ where $q^{d}\in \mathcal Q_0^d$. That is, $q_1=0$ leads to the left sub-tree followed by any sequence of length $d$ that ends with $q_{d}=0$. The individual value function for such sequences is 
    \begin{align}\label{eq:proof_bibala1}
        V^0(0q^{d})&= \frac1{1-r_1} + f_d(y),
    \end{align}
    where we used the facts that the individual value function of the left sub-tree for sequences with $q_{d}=0$ is $f_d(y)$, and that the immediate loss is $\rfrac{r_1}$. Similarly, the individual value functions for the other sequences are 
    \begin{align}\label{eq:proof_bibala2}
        V^1(0q^d)&= y, \ \ \ \ \ \ \ \ \ \ \ \ \ \forall q^d\in \mathcal Q_1^d\nonumber\\
        V^0(1q^d)&= x, \ \ \ \ \ \ \ \ \ \ \ \ \  \forall q^d\in \mathcal Q_0^d \nonumber\\
        V^1(1q^d)&= \frac1{r_1} + f_d(x), \ \ \forall q^d\in \mathcal Q_1^d.
    \end{align}
    Recall that to bi-balance the constructed tree, we should require $V^i(0q^d) = V^i(1q^d)$ for $i=0,1$ (and all $q^d\in\mathcal Q_i^d$). That is, we equate the loss of all paths leading to the same winning team. Using \eqref{eq:proof_bibala1}-\eqref{eq:proof_bibala2}, these two equations can be written as
    \begin{align}\label{eq:proof_bibalan}
        x&= \frac1{1-r_1} + f_d(y) \nonumber\\
        y&= \frac1{r_1} + f_d(x),
    \end{align}
    and can be solved in terms of $x$ as
    \begin{align}\label{eq:proof_bibalan_y}
        y &= \frac{ (d +1 )(x - d)}{ x - ( d + 1)}\nonumber\\
            &= f_{d + 1}(x).
    \end{align}
    This shows that the value function at the new root is $(x,f_{d+1}(x))$. 
    
    It remains to determine the constraints on the value of $x$ such that $r_1 \in (0,1)$.  By \eqref{eq:proof_bibalan}-\eqref{eq:proof_bibalan_y},
    \begin{align}
        1 - r_1 &= \frac1{x - f_{d}(f_{d+1}(x))}\\
                &= \frac{x}{(x-d)^2 + d},
    \end{align}
which is greater than $0$ only if $x>0$. We can also arrange the equation as 
\begin{align}
    1 - r_1 &= \frac{ (x-d) + d}{(x-d)^2+d},
\end{align}
which is smaller than $1$ iff $x>d+1$ or $x-d<0$. Combined with the induction hypothesis, $x>d$, we conclude that for all $x>d+1$, the new odd at the root satisfies $1 - r_1 = \frac1{x - f_{d}(f_{d+1}(x))} \in (0,1)$. We remark that the constructed strategy inherits the strategy at subsequent steps from the sub-trees that are assumed to exist by the induction hypothesis. It can be easily verified that $y = f_{d+1}(x) > d+1$, and we also showed the constraint $x > d+1$. These two conditions guarantee that the strategies at the left and right sub-trees are feasible.
\end{enumerate}
Claim \ref{claim:subtrees} shows the converse: any bi-balanced tree of depth $d+1$ can be constructed by connecting two bi-balanced trees, each of depth $d$. As the sub-trees used for the construction are arbitrary, our construction characterizes all bi-balanced trees of depth $d+1$.
\end{proof}

\subsection{The optimal strategy}\label{subsec:opt_stra}
In this section we utilize Theorem \ref{th:bi-balanced} on the structure of the value function at the root to derive a strategy $\{r_t\}_{t=1}^T$ for the bi-balancing problem. The main idea is that the individual value function at the root can be used to compute the individual value function on any node of the tree in a forward manner. Let $V^i_{q^j}(q_{j+1}^{T})$ denote the value function at the root of the sub-tree whose prefix is $q^j$, and the future bets sequence is $q_{j+1}^{T}$. Throughout this section, we assume that trees are bi-balanced so the dependence on future bets is omitted. The following theorem describes a forward sequential algorithm that finds the individual value functions and the optimal odds as a function of the input $q_1,q_2,\dots$. 
\begin{theorem}\label{th:main_strategy}
    Consider a bi-balanced tree of depth $T$ whose value function at the root is $(V^0,V^1) = (x,f_T(x))$. For any sequence $q_1q_2\dots$, the individual value function along this sequence can be computed recursively as 
    \begin{align}\label{eq:opt_strategy_recur}
        (V_{q^t}^0,V_{q^t}^1) &=
        \begin{cases}
	   (f_{T-t}(V_{q^{t-1}}^1),V_{q^{t-1}}^1), & \text{if $q_{t}=0$}\\
          (V_{q^{t-1}}^0,f_{T-t}(V_{q^{t-1}}^0)), & \text{if $ q_{t}=1$}
	\end{cases}
\end{align}
with the initial condition $(V^0,V^1) = (x,f_T(x))$. The subsequent odd is then equal to 
\begin{align}\label{eq:th_opt_strategy_biblan}
    r_t(q^{t-1}) &= \frac1{V_{q^{t-1}}^1 - V_{q^{t-1}1}^1}.
\end{align}
\end{theorem}
The recursive computation of the individual value functions is the basis for Algorithm \ref{alg:bets_for_bin_2}. Theorem \ref{th:main_strategy} shows that we can update the individual value function in a forward manner so that we only need $t$ computations to execute a strategy up to time $t$. This is a significant reduction in the complexity since in general MDPs (and vector games), where the value function is updated recursively but in a backward fashion. The reduced complexity is due to the fact that, for the online bookmaking problem, the vector of future losses is changing in one of its coordinates at each step. 

\begin{proof}[Proof of Theorem \ref{th:main_strategy}]
    We use the fact that a new bet $q_t\in\{0,1\}$ changes the individual value function only in one of its coordinates. In particular, 
    \begin{align}\label{eq:invariant}
        V_{q^{t-1}0}^1 &= V_{q^{t-1}}^1 \nonumber\\
        V_{q^{t-1}1}^0 &= V_{q^{t-1}}^0.
    \end{align}
 The recursive relation in \eqref{eq:opt_strategy_recur} is completed by Theorem~\ref{th:bi-balanced} showing that the other coordinate of the individual value function can be computed by applying $f_{T-t}(\cdot)$ on the invariant coordinate in \eqref{eq:invariant}. The equation for computing the odd in \eqref{eq:th_opt_strategy_biblan} follows from the relation 
    \begin{align}\label{eq:th_strategy_action}
    V_{q^{t-1}}^1 &= \frac1{r_t} + V_{q^{t-1}1}^1.
    \end{align}
\end{proof}

\subsection{Proof of Theorem \ref{th:main} and Claim \ref{claim:equalizing}}
\begin{proof}[Proof of Theorem \ref{th:main}]
  The online bookmaking problem is a special case of the bi-balancing problem with the constraint $V^0(q^{T-1}0) = V^1(q^{T-1}1)$ for some $q^{T-1}$ (which implies that this holds for all $q^{T-1}\in\{0,1\}^{T-1}$ since the tree is bi-balanced). By Theorem \ref{th:bi-balanced}, this translates to equating the two coordinates of the individual value function at the root, that is,
  \begin{align}
      x&= f_T(x),
  \end{align}
  which simplifies to $x = T \pm \sqrt{T}$. By the range limit, $x > T$, the optimal house loss is $L^*_T = T + \sqrt{T}$.

We proceed to prove the optimality of Algorithm \ref{alg:bets_for_bin_2} for decisive gamblers, i.e., $q^T\in\{0,1\}^T$. The algorithm is essentially a memory-efficient implementation of Theorem \ref{th:main_strategy}. We interpret the variables $a,b,b^+$ that are used in Algorithm \ref{alg:bets_for_bin_2}. The state of Algorithm \ref{alg:bets_for_bin_2}, denoted by $(a,b)$, corresponds to the individual value function $(V^0_{q^{t-1}},V^1_{q^{t-1}})$ after observing the sequence $q^{t-1}$. By \eqref{eq:opt_strategy_recur}, the update of the individual value function is only a function of the previous individual value function (at the invariant coordinate) and the new bet $q_t$. The state update in Algorithm \ref{alg:bets_for_bin_2} is precisely \eqref{eq:opt_strategy_recur}. The variable $b^+$ in Algorithm~\ref{alg:bets_for_bin_2} corresponds to the individual value function $V^1_{q^{t-1}1}$, that is, we hypothetically assume that $q_t=1$. This is utilized to compute the optimal odd in \eqref{eq:th_strategy_action}. By \eqref{eq:th_strategy_action}, first odd is equal to 
\begin{align}
    r_1&= \frac1{V_\emptyset^1 - V^1_1} = 1\frac1{T+\sqrt{T} - f_{T-1}(T+\sqrt{T})} = 1/2.
\end{align}

% \OrComm{Algorithm 1 initializes $r_1=1/2$. It is not clear from the proof why this initialization is good/needed. We need to explain that $V_{\emptyset}^1=T+\sqrt{T}$ and $V_1^1=f_{T-1}(T+\sqrt{T})=T+\sqrt{T}-2$, so that $r_1=\frac{1}{V_{\emptyset}^1-V_1^1}=1/2$.} 

Finally, by Theorem \ref{th:discrete_is_opt}, the house strategy's loss for the expected skeleton strategy~\eqref{eq:strategy_continuous} with respect to the optimal skeleton strategy in Algorithm \ref{alg:bets_for_bin_2} is upper bounded by the optimal house loss.
\end{proof}

\begin{proof}[Proof of Claim \ref{claim:equalizing}]
We show that if a strategy $\{r_t\}$ does not induce an equal loss, i.e., $g(\{r_t\},q^T)$ is the same for all $q^T$, it is not optimal. The key idea is to perturb this strategy to obtain one that achieves a smaller maximum among the losses. For $\{r_t\}$, let $M = \max_{b^T} g(\{r_t\}, b^T)$, and define $\mathcal{M} = \{b^T \in \{0,1\}^T : g(\{r_t\}, b^T) = M\}$. We partition the sequences in $\mathcal{M}$ by identifying the shortest  prefixes such that all subsequent suffixes remain in $\mathcal{M}$. Specifically, a subset of $\mathcal M$ is defined by a prefix $b^i$ and contains $b^i q^{T-i}$ for all $q^{T-i}$, while the latter property does not hold for the prefix $b^{i-1}$. We will now present a procedure to decrease the loss of every sequence in such subsets below $M$. We start by applying the procedure on subsets that have the longest prefix, denoted here by $b^i$, and then apply the procedure on prefixes with shorter lengths. 

The procedure is to \emph{perturb} $r_{i} := \phi_{i}(b^{i-1})$ and subsequent strategies along the sequence $b^i1^{T-i}$, and analyze the effect on the losses (for all binary sequences). In particular, we increase $r_{i}$ and decrease $r_{i+j}:=\phi_{i+j}(b^{i-1}1^j)$ for $j=1,\dots,T-i$. The perturbation clearly has no effect on the losses of sequences whose prefix is not $b^{i-1}$.
    
The losses for sequences with prefix $b^{i-1}$ can be divided into four cases based on the values of $b_i,b_T$. Without loss of generality, assume $b_i=1$. By our partition, all sequences with prefix $b^i$ have a identical maximal loss,
\begin{align}\label{eq:proof_claim2_M}
        M = g(\{r_t\},q^T=b^{i-1}1q_{i+1}^{T-1}0) &= V^0(b^{i-1}) + V_{b^{i-1} 1}^0(q_{i+1}^{T-1}0), \ \ \ \ \ \ \ \ \ \  \forall q_{i+1}^{T-1} \nonumber\\
        M = g(\{r_t\},q^T=b^{i-1}1q_{i+1}^{T-1}1) &= V^1(b^{i-1}) + \frac1{r_{i}} + V_{b^{i-1} 1}^1(q_{i+1}^{T-1}1), \ \ \ \ \forall q_{i+1}^{T-1}.
    \end{align}
    % where we used Claim \ref{claim:winning_horse} to determine the loss function based on $q_T$.\OrComm{I'd delete the sentence ``here we used Claim \ref{claim:winning_horse} to determine the loss function based on $q_T$.''} 
    Also, recall that $V_{b^i}^i(b_{i+1}^T)$ is the value function starting at the sub-tree whose prefix is $b^i$, and its future bets sequence is $b_{i+1}^T$.

    Recall our assumption that $b^i$ is the longest prefix among the subsets of $\mathcal M$. Thus, each sequence with prefix $b^{i-1}0 \neq b^i$ has a loss smaller than $M$, and can be written as
 \begin{align}\label{eq:proof_claim2_m}
        M>g(\{r_t\},q^T=b^{i-1}0q_{i+1}^{T-1}0) &= V^0(b^{i-1}) + \frac1{1 - r_i} + V_{b^{i-1}0}^0(q_{i+1}^{T-1}0), \ \ \ \forall q_{i+1}^{T-1} \nonumber\\
        M>g(\{r_t\},q^T=b^{i-1}0q_{i+1}^{T-1}1) &= V^1(b^{i-1}) + V_{b^{i-1}0}^1(q_{i+1}^{T-1}1) , \ \ \ \ \ \ \ \ \ \ \forall q_{i+1}^{T-1}.
    \end{align}
Our objective is to guarantee that the losses of all sequences with prefix $b^{i-1}$, detailed in \eqref{eq:proof_claim2_M}-\eqref{eq:proof_claim2_m}, is smaller than $M$. The continuity of the functions $\frac1{r},\frac1{1-r}$ implies that $r_i$ can be increased such that the loss of sequences of the form $b^{i-1} 1 q_{i+1}^{T-1}1$ decreases below $M$, while the losses of sequences $b^{i-1} 0 q_{i+1}^{T-1}0$ increase but do not exceed $M$. We are then left with sequences of the form $b^{i-1} 1 q_{i+1}^{T-1}0$ whose losses are not effected by $r_{i}$ and therefore are still equal to $M$. By decreasing $r_{i+j}$, for all $j>0$, the losses of sequences of the form $b^{i-1} 1 q_{i+1}^{T-1}0$ decrease below $M$. However, by doing so, we also increase the losses of sequences of the form $b^{i-1} 1 q_{i+1}^{T-1}1$ (whose losses are smaller than $M$). Again, by continuity, we can guarantee that the perturbations are small enough such that these losses do not exceed $M$.

At the end of the perturbation procedure described above, all sequences with prefix $b^{i-1}$ have losses that are smaller than $M$ and, therefore, can be eliminated from $\mathcal M$. We repeat the procedure above for any subset of sequences whose prefix length is $i$. We then apply the procedure for prefixes of length $i-1$ and so on until all sequences in $\mathcal M$ have a loss smaller than $M$. 
\end{proof}

\section{Monte-Carlo-Based Approximation of the Optimal Algorithm}
\label{sec:montecarlo}

In this section, we discuss complexity aspects of implementing the optimal house strategy to calculate $r_t$ and provide an efficient algorithm to closely approximate $r_t$ for general gamblers (i.e. non-decisive gamblers who can potentially choose $q_t \in [0,1]$). 

First, we note that if the gamblers are decisive, Algorithm~\ref{alg:bets_for_bin_2} which computes the optimal strategy has a linear (in $T$) runtime---this follows since at each time step simple computations are executed to update the state $(a,b)$ and compute $r_{t}$ as a function of the previous state, $t$, and the gambler's previous bet $q_{t-1}$. 

For general non-decisive gamblers, computing the odds $\overline{r}^{\texttt{ALG}}_t(q^{t-1})=\EE[{r}^{\texttt{ALG}}_t(X^{t-1})]$, as defined in~\eqref{eq:strategy_continuous} can take exponential time as discussed in Section~\ref{subsec:algcont}.
% . This is because in this case
% \begin{align} \label{eq:OptActionGeneral}
% \bar{r}^{\texttt{ALG}}_t = \EE[\phi_t^{\texttt{ALG}}(X^{t-1})] 
% \end{align}
% \os{The notation $r^{\texttt{ALG}}_t$ may confuse later in this section, I would add a bar over it to emphasize it is produced from the EV}\OrComm{I changed the notation as you suggested} where $X_i\sim \text{Ber}(q_i)$, so that 
% \[
% \bar{r}^{\texttt{ALG}}_t = \sum_{x^{t-1}\in\{0,1\}^{t-1}} \phi_t^{\texttt{ALG}}(x^{t-1}) \prod_{i=1}^{t-1} q_i^{x_i}(1-q_i)^{1-x_i}
% \]
% requiring an exponentially large number of computations per time step.  
In order to achieve an efficient algorithm to calculate $\overline{r}^{\texttt{ALG}}_t$ in this general case, we propose to evaluate $\EE[r_t^{\texttt{ALG}}(X^{t-1})]$ via a Monte Carlo simulation in lieu of an exact computation. If $X_{(1)}^{t-1},\dotsc, X_{(N)}^{t-1}$ are $N$ independent copies of $X^{t-1}$, then 
\begin{align} \label{eq:MonteCarloSum}
\frac{1}{N}\sum_{j=1}^N r_t^{\texttt{ALG}}(X_{(j)}^{t-1}) \approx  \EE[r_t^{\texttt{ALG}}(X^{t-1})].
\end{align}

The Monte Carlo method based algorithm (Algorithm~\ref{alg:MonteCarlo}) essentially runs $N$ parallel copies of Algorithm~\ref{alg:bets_for_bin_2}, keeping track of $N$ states $(a_{(j)},b_{(j)})_{j=1}^N$. Upon receiving the gambler's bet $q_t \in [0,1]$ the algorithm generates $X_{(j)t} \sim \text{Ber}(q_t)$ , updates the state $(a_{(j)},b_{(j)})$ and generates an action for the house as done in Algorithm~\ref{alg:bets_for_bin_2}; the $N$ copies of this action are then averaged to approximate the odds $\overline{r}^{\texttt{ALG}}_t(q^{t-1}) = \EE[r_t^{\texttt{ALG}}(X^{t-1})]$ as in~\eqref{eq:MonteCarloSum}. Algorithm~\ref{alg:MonteCarlo} has two (nested) loops of size $N$ and $T$ with a constant number of operations in each so that the total time complexity is $\mathcal{O}(NT)$. 

\begin{algorithm}[!h]
\caption{Monte Carlo Based Efficient Strategy}
\label{alg:MonteCarlo}
\begin{algorithmic}
\State \textbf{Inputs:} $T$ (Total rounds), $q_1,\ldots,q_{T-1}$ (Gambler's bets obtained sequentially)
\State \textbf{Output:} $r_1,\ldots,r_t$ (House strategy, $r_t$ is output after obtaining $q_1,\ldots,q_{t-1}$)
\State \textbf{Initialization:} $a_{(j)} = b_{(j)} = T+ \sqrt{T}, \text{ } j \in [N]$ 
\State $r_1\gets \frac{1}{2}$
\State Output $r_{1}$
\For{$t=1:T-1$}
\State $r_{t+1} \gets 0$
\State $d \gets T - t$
\For{$j = 1:N$}
\State $X \sim \text{Ber}(q_t)$
\If{$X = 0$} 
    \State $a_{(j)} \gets d \frac{b_{(j)} - (d-1)}{b_{(j)} - d}$ 
\ElsIf{$X = 1$}
    \State $b_{(j)} \gets d \frac{a_{(j)} - (d-1)}{a_{(j)} - d}$ 
\EndIf
\State $b^+_{(j)} \gets (d-1) \frac{a_{(j)} - (d-2)}{a_{(j)}-(d-1)}$ 
\State $r_{t+1} \gets r_{t+1} + \frac{1}{N} \cdot \frac1{b_{(j)}-b^+_{(j)}}$
\EndFor
\State Output $r_{t+1}$
\EndFor
\end{algorithmic}
\end{algorithm}

As $N$ increases, the approximation calculated by Algorithm~\ref{alg:MonteCarlo} converges rapidly to the desired odds $\{\overline{r}^{\texttt{ALG}}_t\}$. This rate of convergence can be quantified as follows.

\begin{proposition} \label{prop:MonteCarloConvergence}
    Let $\overline{r}^{\texttt{ALG}}_t=\overline{r}^{\texttt{ALG}}_t(q^{t-1}) = \EE[r_t^{\texttt{ALG}}(X^{t-1})]$, and  $\widehat{r}^N_t=\widehat{r}^N_t(q^{t-1}) :=  \frac{1}{N}\sum_{j=1}^N r_t^{\texttt{ALG}}(X_{(j)}^{t-1})$ be its approximation computed by Algorithm~\ref{alg:MonteCarlo}. Then for any $\epsilon > 0$, and $q^T\in[0,1]^T$, we have that $\Pr(|\widehat{r}^N_t-\overline{r}^{\texttt{ALG}}_t| \ge \epsilon) \le 2\exp\left(-2N\epsilon^2 \right)$ and subsequently $\Pr(|\widehat{r}^N_t-\overline{r}^{\texttt{ALG}}_t| \ge \epsilon \text{ for any } t \in [T]) \le 2T\exp\left(-2N\epsilon^2 \right)$
\end{proposition}

\begin{proof}
    Since $\widehat{r}^N_t$ is a sum of $N$ independent and identically distributed random variables $r_t^{\texttt{ALG}}(X_{(1)}^{t-1}),\dotsc,r_t^{\texttt{ALG}}(X_{(N)}^{t-1})$ each bounded in $[0,1]$ with mean $\overline{r}^{\texttt{ALG}}_t(q^{t-1})$, the first assertion follows by the Hoeffding inequality. The second assertion follows via the union bound. 
\end{proof}

Proposition~\ref{prop:MonteCarloConvergence} implies that in order to achieve $\Pr(|\widehat{r}^N_t-\overline{r}^{\texttt{ALG}}_t| \ge \epsilon \text{ for any } t \in [T]) \le \delta$, choosing any $N \ge \frac{1}{2\epsilon^2} \log \left(\frac{2T}{\delta}\right)$ suffices. In order to choose an appropriate value of $\epsilon$, we consider the effect of choosing the odds $\widehat{r}^N_t$ instead of the desired odds $\overline{r}_t^{\texttt{ALG}}$. In particular, $\epsilon$ should be small enough so that the worst-case loss
\begin{align}\label{eq:MonteCarloApproxLoss}
    \widehat{L}^N_T := \max_{q^T} \max\left\{\sum_{t=1}^T \frac{q_t}{\widehat{r}^N_t}, \sum_{t=1}^T \frac{1-q_t}{1-\widehat{r}^N_t}\right\}
\end{align}
obtained by playing the (suboptimal) action $\widehat{r}^N_t$ is close to $L^*_T$. To do this, we utilize the following. 
\begin{claim} \label{claim:MinProb}
    For fixed horizon size $T$, and any $q^T\in[0,1]^T$ we have $\frac{1}{T+\sqrt{T}} \le \overline{r}^{\texttt{ALG}}_t(q^{t-1}) \le 1-\frac{1}{T+\sqrt{T}}$ for all $1\leq t\leq T$. 
\end{claim}

\begin{proof}
This follows from the fact that the strategy $\{\overline{r}^{\texttt{ALG}}_t\}$ attains the optimal house loss $L_T^*=T+\sqrt{T}$. Indeed, assume for contradiction that there is some $q^{t-1}$ for which $\overline{r}^{\texttt{ALG}}_t(q^{t-1})<\frac{1}{T+\sqrt{T}}$. Then, the individual game loss for any $q^T$ beginning with $q^{t-1}1$ is at least $\frac{1}{\overline{r}^{\texttt{ALG}}_t(q^{t-1})}>T+\sqrt{T}=L^*_T$, in contradiction to the fact that $\{\overline{r}^{\texttt{ALG}}_t\}$ attains the optimal house loss $L_T^*$. The upper bound $\overline{r}^{\texttt{ALG}}_t(q^{t-1}) \le 1-\frac{1}{T+\sqrt{T}}$ is proved similarly.
\end{proof}

\begin{lemma} \label{lem:SuboptimalLoss}
Let $\delta\in(0,1)$, assume $\epsilon \le \frac{1}{2(T+\sqrt{T})}$ and  $N \ge \frac{1}{2\epsilon^2} \log \left(\frac{2T}{\delta}\right)$. Let $\{\overline{r}^{\texttt{ALG}}\}$ be the house strategy computed by Algorithm~\ref{alg:bets_for_bin_2} and~\eqref{eq:strategy_continuous}, and $\{\widehat{r}^{N}\}$ be the house strategy computed by Algorithm~\ref{alg:MonteCarlo}. Then, with probability at least $1-\delta$, for any $q^T\in[0,1]^T$ the corresponding individual game losses satisfy
\begin{align}
\|\varphi_T\left(\{\widehat{r}^{N}\},q^T \right)\|_{\infty}\leq \left( 1+2\epsilon(T+\sqrt{T})\right)\|\varphi_T\left(\{\overline{r}^{\texttt{ALG}}\},q^T \right)\|_{\infty}\leq \left( 1+2\epsilon(T+\sqrt{T})\right) L_T^*. 
\end{align}
\end{lemma}

\begin{proof}
Since $N \ge \frac{1}{2\epsilon^2} \log \left(\frac{2T}{\delta}\right)$, Proposition~\ref{prop:MonteCarloConvergence} ensures that the event $|\widehat{r}^N_t - \overline{r}^{\texttt{ALG}}_t| \le \epsilon$ for all $t \in [T]$ occurs with probability ta least $1-\delta$. If this event occurs, then
\[
    \frac{q_t}{\widehat{r}^N_T} \le \frac{q_t}{\overline{r}^{\texttt{ALG}}_t-\epsilon} \le \frac{q_t}{\overline{r}^{\texttt{ALG}}_t}\cdot \frac{1}{1-\epsilon(T+\sqrt{T})}
\]
and similarly, 
\[
\frac{1-q_t}{1-\widehat{r}^N_T} \le \frac{1-q_t}{1-\overline{r}^{\texttt{ALG}}_t}\cdot \frac{1}{1-\epsilon(T+\sqrt{T})}.
\]
where both of these use Claim~\ref{claim:MinProb} to bound $\overline{r}^{\texttt{ALG}}_t$. Putting these two together implies that for any $q^T \in [0,1]^T$, with probability $1-\delta$
\begin{align} \label{eq:ApproxLbound}
   \|\varphi_T\left(\{\widehat{r}^{N}\},q^T \right)\|_{\infty}&= \max\left\{\sum_{t=1}^T \frac{q_t}{\widehat{r}^N_T(q^{t-1})}, \sum_{t=1}^T \frac{1-q_t}{1-\widehat{r}^N_T(q^{t-1})}\right\} \\
   &\le \frac{1}{1-\epsilon(T+\sqrt{T})}\max\left\{\sum_{t=1}^T \frac{q_t}{\overline{r}^{\texttt{ALG}}_t(q^{t-1})}, \sum_{t=1}^T \frac{1-q_t}{1-\overline{r}^{\texttt{ALG}}_t(q^{t-
   1})}\right\}\\
   &\le \left(1+2\epsilon(T+\sqrt{T})\right)\|\varphi_T\left(\{\overline{r}^{\texttt{ALG}}\},q^T \right)\|_{\infty} \label{eq:1over1m}\\
   &\le \left(1+2\epsilon(T+\sqrt{T})\right) L^*_T,\label{eq:LTUB}
\end{align}
where~\eqref{eq:1over1m} uses that $\frac{1}{1-x} \le 1+2x$ for $0\le x \le 1/2$, and~\eqref{eq:LTUB} the fact that $\max_{q^T\in[0,1]^T}\|\varphi_T\left(\{\overline{r}^{\texttt{ALG}}\},q^T \right)\|_{\infty}=L_T^*$ from the fact that the strategy $\{\overline{r}^{\texttt{ALG}}\}$ attains the optimal house loss.
\end{proof}

In order to get $1+2\epsilon(T+\sqrt{T}) \to 1$, we must choose $N$ large enough so that the corresponding $\epsilon = o\left(\frac{1}{T}\right)$. For example,  $\epsilon = \frac{1}{T^{1+\alpha}}$ for any $\alpha > 0$ suffices. 
Recalling that the time complexity of Algorithm~\ref{alg:MonteCarlo} is $\mathcal{O}(NT)$ and that in order to achieve an $\epsilon$ (additive) approximation to $\overline{r}^{\texttt{ALG}}_t$ with probability $1-\delta$ requires $N \ge \frac{T}{2\epsilon^2} \log \left(\frac{2T}{\delta}\right)$, we get that for choice of $\epsilon = \frac{1}{T^{1+\alpha}}$ 
the time complexity of Algorithm~\ref{alg:MonteCarlo} is $\mathcal{O}\left(T^{3+2\alpha}\log \left(\frac{T}{\delta}\right)\right)$. Even for fairly small values of $\delta$, this  is far smaller than the $\mathcal{O}\left(T2^T\right)$ complexity of computing $\overline{r}^{\texttt{ALG}}_t$ exactly.

\section{Sub-optimal Blackwell-Inspired Algorithm}
\label{sec:Blackwell}

The purpose of this section is to show that house strategies for the online bookmaking game, whose amortized loss tends to $1$ as $T\to\infty$, can also be derived through standard approaches. However, these strategies are sub-optimal, and do not even attain a $O(T^{-1/2})$ rate of convergence. We only consider the binary case, but the algorithm below can be extended to $m>2$ as well.

\medskip

We follow  Blackwell's algorithm~\cite{blackwell1956analog} (see also~\cite{FarinaLectureNotes}), to derive a sub-optimal algorithm for the optimal bookmaking problem. Since our loss functions are unbounded in general, we cannot apply Blackwell's algorithm as-is in order to approach an $\ell_{\infty}$ ball, and some adaptation is required. Because of this adaptation, we can only achieve a regret of $O(T^{-1/4})$, rather than the $O(T^{-1/2})$ rate of approachability that Blackwell's algorithm provides for bounded loss functions.

\begin{figure}[h]
    \centering

\begin{tikzpicture}[scale=3]

% Set the parameters
\def\DeltaA{1.2}
\def\K{2}

% Calculate coordinates for a_1, a_2, a_3
\coordinate (a1) at ({1 + \K*\DeltaA*(1/\DeltaA)}, {1 + \K*\DeltaA*(1 - 1/\DeltaA)});
\coordinate (a2) at ({1 + \K*(1 - \DeltaA)}, {1 + \K*1});
\coordinate (a3) at ({\DeltaA + (\K + 1 - \DeltaA)*\DeltaA*(1/\DeltaA)}, {(\K + 1 - \DeltaA)*\DeltaA*(1 - 1/\DeltaA)});

% Calculate coordinates for b_1, b_2, b_3
\coordinate (b1) at ({1 + \K*\DeltaA*(1 - 1/\DeltaA)}, {1 + \K*\DeltaA*(1/\DeltaA)});
\coordinate (b2) at ({1 + \K*1}, {1 + \K*(1 - \DeltaA)});
\coordinate (b3) at ({(\K + 1 - \DeltaA)*\DeltaA*(1 - 1/\DeltaA)}, {\DeltaA + (\K + 1 - \DeltaA)*\DeltaA*(1/\DeltaA)});

% Axis
\draw[->] (-0.3,0) -- (1+\K,0) node[right] {$x_1$};
\draw[->] (0,-0.3) -- (0,1+\K) node[above] {$x_2$};

% Red vectors for lambda_2 and lambda_2^perp
\draw[->, thick, red] (1,1) -- (a1) node[midway, above right] {$\lambda_2$};
\draw[->, thick, red] (\DeltaA,0) -- (a2) node[midway, below left] {$\lambda_2^\perp$};

% Green vectors for lambda_1 and lambda_1^perp
\draw[->, thick, green!20!black] (1,1) -- (b1) node[midway, above left] {$\lambda_1$};
\draw[->, thick, green!50!black] (0,\DeltaA) -- (b2) node[midway, above right] {$\lambda_1^\perp$};

% Polytope (LightGreen A_1^-)
\filldraw[fill=green!10, draw=green!50!black, opacity=0.5] (0,\DeltaA) -- (0,{1+\K}) -- (b3) -- (0,\DeltaA) -- cycle;

% Polytope (darkGreen A_1^+)
\filldraw[fill=green!40, draw=green!50!black, opacity=0.5] (0,\DeltaA) -- (b3)-- (b1)--(1,1) -- (0,\DeltaA) -- cycle;

% Polytope (LightRed B_2^-)
\filldraw[fill=red!10, draw=red!50!black,  opacity=0.5] (\DeltaA,0) -- ({1+\K},0) -- (a3) -- (\DeltaA,0) -- cycle;

% Polytope (darkRed B_2^+)
 \filldraw[fill=red!40, draw=red!50!black,  opacity=0.5] (\DeltaA,0) -- (a3)-- (a1)--(1,1) -- (\DeltaA,0) -- cycle;

% Polytope (lightGrey A_3)
 \filldraw[fill=black!20, draw=black!50!black,  opacity=0.5] (1,1)-- (a1)--({1+\K},{1+\K}) -- (b1)--(1,1) -- cycle;

% Polytope (Purple S)
 \filldraw[fill=blue!20, draw=blue!50!black,  opacity=0.5] (0,0)-- (0,\DeltaA)--(1,1) -- (\DeltaA,0)--(0,0) -- cycle;

% Labels for polytope
 \node at (0,\DeltaA) [circle,fill=black,inner sep=1pt] {} ;
 \node at (0,\DeltaA) [left] {$(0,\Delta)$} ;

 \node at (\DeltaA,0) [circle,fill=black,inner sep=1pt] {} ;
 \node at (\DeltaA,0) [below] {$(\Delta,0)$} ;

 \node at (1,1) [circle,fill=black,inner sep=1pt] {} ;
 \node at (1,1) [below left] {$(1,1)$} ;

  \node at (0.5,0.5)  {$\m{S}$} ;
  \node at (1.5,1.5)  {$\m{A}_3$};
  \node at (b3) [below left] {$\m{A}_1^{-}$};
  \node at (b3) [below right] {$\m{A}_1^{+}$};
  \node at (a3) [below left] {$\m{A}_2^{-}$};
  \node at (a3) [above left] {$\m{A}_2^{+}$};
%
% \node at (0,{1+\K}) [circle,fill=green,inner sep=1pt] {};
% \node at (b3) [circle,fill=green,inner sep=1pt] {};

\end{tikzpicture}

	\caption{Illustration of the regions defined in~\eqref{eq:BlackwellRegions} and in~\eqref{eq:BlackwellRegionsRefinement}. Here $\m{A}_1=\m{A}_1^{-}\cup\m{A}_1^{+}$ and $\m{A}_2=\m{A}_2^{-}\cup\m{A}_2^{+}$.}
	\label{fig:BlackwellRegions}
\end{figure}
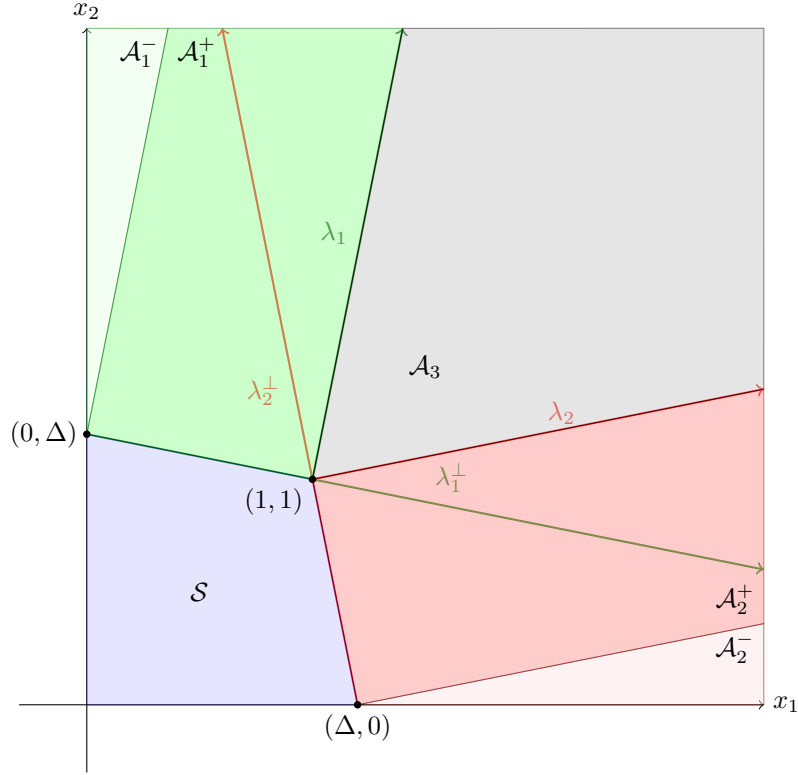

While in previous sections it was convenient to think of probability vectors in the simplex $\Delta_2$ as points $r_t,q_t\in[0,1]$, in this section $r_t,q_t$ will refer to two-dimensional vectors in the simplex. Recall that for two vectors $r=(r(0),r(1))$ and $q=(q(0),q(1))$ in the simplex $\Delta_2$ our loss function is
\begin{align}
f(r,q)=\left(\frac{q(0)}{r(0)},\frac{q(1)}{r(1)}\right)^\tr.    
\end{align}
For some $1<\Delta<2$ to be specified later, define the vectors
\begin{align}
\lambda_1&=\left(1-\frac{1}{\Delta},\frac{1}{\Delta}\right)^\tr,~~\lambda_1^{\perp}=\left(1,1-\Delta\right)^\tr\label{eq:lambda1}\\
\lambda_2&=\left(\frac{1}{\Delta},1-\frac{1}{\Delta}\right)^\tr,~~\lambda_2^{\perp}=\left(1-\Delta,1\right)^\tr.\label{eq:lambda2}
\end{align}
Define the sets $\m{S},\m{A}_1,\m{A}_2,\m{A}_3\subset [0,\infty)^2$ as
\begin{align}
\m{S}&=\left\{x\in [0,\infty)^2~:~(x-(1,1)^\tr)^\tr\lambda_1 \leq 0,(x-(1,1)^\tr)^\tr\lambda_2 \leq 0 \right\} \nonumber\\
\m{A}_1&=\left\{x\in [0,\infty)^2~:~(x-(1,1)^\tr)^\tr\lambda_1>0,(x-(1,1)^\tr)^\tr\lambda_1^{\perp} \leq 0 \right\}\nonumber\\
\m{A}_2&=\left\{x\in [0,\infty)^2~:~ (x-(1,1)^\tr)^\tr\lambda_2>0,(x-(1,1)^\tr)^\tr\lambda_2^{\perp} \leq 0 \right\}\nonumber\\
\m{A}_3&=\left\{x\in [0,\infty)^2~:~(x-(1,1)^\tr)^\tr\lambda_1^{\perp} >0,(x-(1,1)^\tr)^\tr\lambda_2^{\perp} > 0 \right\}
\label{eq:BlackwellRegions}
\end{align}
See Figure~\ref{fig:BlackwellRegions}.
% \begin{figure}[t]
% 	\includegraphics[width=0.9\textwidth]{BlackwellAlgorithmRegions.png}
% 	\centering
% 	\caption{Illustration of the regions defined in~\eqref{eq:BlackwellRegions}}
% 	\label{fig:BlackwellRegions}
% \end{figure}

The algorithm is as follows. Let $\bar{\vphi}_{0}=(0,0)^\tr$. For any $t=1,2,\ldots$ let 
\begin{align}
\bar{\vphi}_t=\frac{1}{t}\sum_{i=1}^t f(r_i,q_i), 
\label{eq:phitdef}
\end{align}
and for any $t=0,\ldots,T-1$ set 
\begin{align}
r_{t+1}=\begin{cases}
\lambda_1 &    \bar{\vphi}_t\in \m{A}_1\\
\lambda_2 &    \bar{\vphi}_t\in \m{A}_2\\
\frac{\bar{\vphi}_t-(1,1)^\tr}{\|\bar{\vphi}_t-(1,1)^\tr\|_1} & \bar{\vphi}_t\in\m{A}_3\\
(\frac{1}{2},\frac{1}{2})^\tr & \bar{\vphi}_t\in S
\end{cases}.
\label{eq:BlackwellUpdateRule}
\end{align}
Note that $r_{t+1}=\phi_{t+1}(q_1,\ldots,q_{t})$ is a deterministic function of the previous points in the simplex chosen by the gambler. Thus, it is a valid strategy for the binary online bookmaking game.

\begin{theorem}\label{th:blackwell}
For any $1<\Delta< 2$, and any $q_1,\ldots,q_t\in\Delta_2$, the algorithm specified by equation~\eqref{eq:BlackwellUpdateRule}, with $\lambda_1$, $\lambda_2$, $\lambda^{\perp}_1$, $\lambda^{\perp}_2$, $\m{A}_1$, $\m{A}_2$, $\m{A}_3$, $\m{S}$, $\bar{\vphi}_t$, defined by equations (\ref{eq:lambda1}-\ref{eq:phitdef}) attains
\begin{align}
\frac{1}{t}\|\vphi_t(\{\phi_t\},q^t)\|_{\infty}=\|\bar{\vphi}_t\|_{\infty}\leq \Delta+\sqrt{\frac{2}{t}}\frac{\Delta^2}{\Delta-1}.
\end{align}
In particular, if we apply the algorithm for $T$ iterations with $\Delta=1+\delta_T$, where
\begin{align}
\delta_T=\frac{\left(\frac{2}{T}\right)^{1/4}}{1-\left(\frac{2}{T}\right)^{1/4}},    
\end{align}
the algorithm attains
\begin{align}
\frac{1}{T}\|\vphi_T(\{\phi_t\},q^T)\|_{\infty}=\|\bar{\vphi}_T\|_{\infty}\leq 1+2\delta_T=1+2\frac{\left(\frac{2}{T}\right)^{1/4}}{1-\left(\frac{2}{T}\right)^{1/4}}.
\end{align}
\end{theorem}

\begin{proof}[Proof of Theorem \ref{th:blackwell}]
For a point $x\in[0,\infty)^2$, and a set $\m{A}\subset\RR^2$ let
\begin{align}
\dist(x,\m{A})=\min_{a\in\m{A}}\|x-a\|    
\end{align}
denote the distance between $x$ and $\m{A}$, and
\begin{align}
\pi_{\m{A}}(x)=\argmin_{a\in \m{A}}\|x-a\|
\end{align}
denote the projection of $x$ on $\m{A}$. Note that $\dist(x,\m{A})=\|x-\pi_{\m{A}}(x)\|$.

Let us define the intervals
\begin{align}
\m{H}_1&=\left\{x\in[0,\infty)^2~:~(x-(1,1)^\tr)^\tr\lambda_1=0 \right\},\\
\m{H}^{-}_1&=\left\{x\in[0,\infty)^2~:~(x-(1,1)^\tr)^\tr\lambda_1=0,(x-(1,1)^\tr)^\tr\lambda^{\perp}_1\in[-1,0] \right\}\\
\m{H}_2&=\left\{x\in[0,\infty)^2~:~(x-(1,1)^\tr)^\tr\lambda_2=0 \right\},\\
\m{H}_2^{-}&=\left\{x\in[0,\infty)^2~:~(x-(1,1)^\tr)^\tr\lambda_2=0,(x-(1,1)^\tr)^\tr\lambda^{\perp}_2\in[-1,0] \right\},
\end{align}
such that the boundary of $\m{S}$ is $\m{H}_1^{-}\cup\m{H}_2^{-}$.
Clearly, for $x\in\m{S}$ we have that $\dist(x,\m{S})=0$ and $\pi_{\m{S}}(x)=x$. For any $x\notin\m{S}$ we have that $\pi_{\m{S}}(x)$ lies on the boundary of $\m{S}$, and therefore
\begin{align}
\dist(x,S)&=\min\{\dist(x,\m{H}^{-}_1),\dist(x,\m{H}^{-}_2)\}\\  
\pi_{\m{S}}(x)&=\argmin_{\pi_{\m{H}_1^{-}}(x),\pi_{\m{H}_2^{-}}(x)}\{\|x-\pi_{\m{H}_1^{-}}(x)\|, \|x-\pi_{\m{H}_2^{-}}(x)\|\}.
\end{align}
With this, it is easy to see (refer to Figure~\ref{fig:BlackwellRegions}) that 
\begin{align}
\pi_{\m{S}}(x)=\begin{cases}
\pi_{\m{H}^{-}_1}(x) & x\in\m{A}_1\\
\pi_{\m{H}^{-}_2}(x) & x\in\m{A}_2\\
\pi_{\m{H}^{-}_1}(x)=\pi_{\m{H}^{-}_2}(x)=(1,1)^\tr & x\in\m{A}_3\\
x & x\in\m{S}
\end{cases}.    
\end{align}
To write $\pi_{\m{S}}(x)$ more explicitly, we define the partition of the sets $\m{A}_1$ and $\m{A}_2$ to
\begin{align}
\m{A}_1^{-}&=\left\{x\in\m{A}_1~:~(x-(1,1)^\tr)^\tr \lambda_1^{\perp}<-1\right\},~\m{A}_1^+=\m{A}_1\setminus\m{A}_1^-,\nonumber\\
\m{A}_2^{-}&=\left\{x\in\m{A}_2~:~(x-(1,1)^\tr)^\tr \lambda_2^{\perp}<-1\right\},~\m{A}_2^+=\m{A}_2\setminus\m{A}_2^-,
\label{eq:BlackwellRegionsRefinement}
\end{align}
such that
\begin{align}
\pi_{\m{S}}(x)=\begin{cases}
(0,\Delta)^\tr=(1,1)^\tr-\lambda_1^{\perp} & x\in\m{A}_1^{-}\\
(1,1)^\tr+((x-(1,1)^\tr)^\tr \lambda_1^{\perp})\cdot \lambda_1^{\perp} & x\in\m{A}^+_1\\
(1,1)^\tr+((x-(1,1)^\tr)^\tr \lambda_2^{\perp})\cdot \lambda_2^{\perp} & x\in\m{A}^+_2\\
(\Delta,0)^\tr=(1,1)^\tr-\lambda_2^{\perp} & x\in\m{A}_2^{-}\\
(1,1)^\tr & x\in\m{A}_3\\
x & x\in\m{S}
\end{cases}.    
\label{eq:projectionexplicit}
\end{align}

Let $\bar{\vphi}_t$ be as defined in~\eqref{eq:phitdef}, and $\eta_t=\pi_S(\bar{\vphi}_t)$. Furthermore, let
\begin{align}
\rho_t=\min_{s\in\m{S}}\|\bar{\vphi}_t-s\|^2=\|\bar{\vphi}_t-\pi_{\m{S}}(\bar{\vphi}_t)\|^2=\|\bar{\vphi}_t-\eta_t\|^2.    
\end{align}
We have 
\begin{align}
\bar{\vphi}_{t+1}=\frac{1}{t+1}\sum_{i=1}^{t+1}f(r_i,q_i)=\frac{t}{t+1}\bar{\vphi}_t+\frac{1}{t+1}f(r_{t+1},q_{t+1}).    
\end{align}
Consequently,
\begin{align}
\rho_{t+1}&=    \min_{s\in\m{S}}\|\bar{\vphi}_{t+1}-s\|^2\nonumber\\
&\leq \|\bar{\vphi}_{t+1}-\eta_t\|^2\\
&=\left\|\frac{t}{t+1}(\bar{\vphi}_t-\eta_t)+\frac{1}{t+1}(f(r_{t+1},q_{t+1})-\eta_t) \right\|^2\\
&=\frac{1}{(t+1)^2}\left(t^2 \rho_t+\|f(r_{t+1},q_{t+1})-\eta_t \|^2+2t(\bar{\vphi}_t-\eta_t)^\tr (f(r_{t+1},q_{t+1})-\eta_t)  \right).
\label{eq:rhoPartial}
\end{align}
We now show that $(\bar{\vphi}_t-\eta_t)^\tr (f(r_{t+1},q_{t+1})-\eta_t)\leq 0$. We first claim that for all $\bar{\vphi}_t\in\m{A}_1$ it holds that $f(r_{t+1},q_{t+1})-\eta_t=\beta_1 \lambda_1^{\perp}$ for some $\beta_1\in\mathbb{R}$. To see this, note that $\eta_t=\pi_{\m{S}}(\bar{\vphi}_t)=(1,1)^\tr+\gamma\lambda_1^\perp$ for all $\bar{\vphi}_t\in\m{A}_1$ for some $\gamma\in[-1,0]$. Thus,
\begin{align}
\lambda_1^\tr\left(f(r_{t+1},q_{t+1})-\eta_t \right)=(\lambda_1(0),\lambda_1(1))\left[\left( \frac{q_{t+1}(0)}{\lambda_1(0)},\frac{q_{t+1}(1)}{\lambda_1(1)}\right)^\tr-(1,1)^\tr-\gamma \lambda_1^{\perp}\right]=0,
\end{align}
where the last equality follows since both $\lambda_1$ and $q_{t+1}$ are probability vectors. We claim further, that for $\bar{\vphi}_t\in\m{A}_1^{-}$, we have that $\beta_1\geq 0$. This follows since both $f(r_{t+1},q_{t+1})$ and $\eta_t$ are on the line $\m{H}_1$, but $\eta_t=(0,\Delta)^\tr$ is the leftmost point on this line. Similarly, for all $\bar{\vphi}_t\in\m{A}_2$ it holds that $f(r_{t+1},q_{t+1})-\eta_t=\beta_2 \lambda_2^{\perp}$, and furthermore, $\beta_2\geq 0$ if $\eta_t\in\m{A}_2^{-}$.

Next, note that (recalling~\eqref{eq:projectionexplicit})
\begin{align}
\frac{\bar{\vphi}_t-\eta_t}{\|\bar{\vphi}_t-\eta_t\|_1}=\begin{cases}
\frac{\bar{\vphi}_t-(0,\Delta)^\tr}{\|\bar{\vphi}_t-(0,\Delta)^\tr\|_1}=a_{1_-}\cdot\lambda_1+b_{1_-}\cdot\lambda_1^{\perp} &    \bar{\vphi}_t\in \m{A}^-_1\\
\lambda_1 &    \bar{\vphi}_t\in \m{A}^+_1\\
\lambda_2 &    \bar{\vphi}_t\in \m{A}^+_2\\
\frac{\bar{\vphi}_t-(\Delta,0)^\tr}{\|\bar{\vphi}_t-(\Delta,0)^\tr\|_1}=a_{2_-}\cdot\lambda_2+b_{2_-}\cdot\lambda_2^{\perp} &    \bar{\vphi}_t\in \m{A}^-_2\\
\frac{\bar{\vphi}_t-(1,1)^\tr}{\|\bar{\vphi}_t-(1,1)^\tr\|_1} & \bar{\vphi}_t\in\m{A}_3\\
(0,0)^\tr & \bar{\vphi}_t\in S
\end{cases},
\end{align}
where $b_{1_-},b_{2_-}\leq 0$ (and also $a_{1_-},a_{2_-}\geq 0$). 
Thus,
\begin{align}
\frac{(\bar{\vphi}_t-\eta_t)^\tr (f(r_{t+1},q_{t+1})-\eta_t)}{\|\bar{\vphi}_t-\eta_t\|_1}&=\begin{cases}
(a_{1_-}\cdot\lambda_1+b_{1_-}\cdot\lambda_1^{\perp})^\tr \beta_1\lambda_1^{\perp} &    \bar{\vphi}_t\in \m{A}^-_1\\
\lambda_1^\tr \beta_1\lambda_1^{\perp} &    \bar{\vphi}_t\in \m{A}^+_1\\
\lambda_2^\tr \beta_2\lambda_2^{\perp} &    \bar{\vphi}_t\in \m{A}^+_2\\
(a_{2_-}\cdot\lambda_2+b_{2_-}\cdot\lambda_2^{\perp})^\tr \beta_2\lambda_2^{\perp} &    \bar{\vphi}_t\in \m{A}^-_2\\
q_{t+1}(1)+q_{t+1}(2)-\frac{\bar{\vphi}_t-(1,1)^\tr}{\|\bar{\vphi}_t-(1,1)^\tr\|_1}(1,1) & \bar{\vphi}_t\in\m{A}_3\\
0 & \bar{\vphi}_t\in S
\end{cases}  \nonumber\\
&=\begin{cases}
 b_{1-}\cdot\beta_1\cdot\|\lambda_1^{\perp}\|^2 &    \bar{\vphi}_t\in \m{A}^-_1\\
0 &    \bar{\vphi}_t\in \m{A}^+_1\\
0 &    \bar{\vphi}_t\in \m{A}^+_2\\
b_{2-}\cdot\beta_2\cdot\|\lambda_2^{\perp}\|^2 &    \bar{\vphi}_t\in \m{A}^-_2\\
0 & \bar{\vphi}_t\in\m{A}_3\\
0 & \bar{\vphi}_t\in S,
\end{cases}
\end{align}
and this is non-positive in all regions $\m{A}^{-}_1,\m{A}^{+}_1,\m{A}^{-}_2,\m{A}^{+}_2,\m{A}_3,\m{S}$.

% and that (recalling~\eqref{eq:BlackwellUpdateRule})
% \begin{align}
% f(r_{t+1},q_{t+1})-\eta_t=\begin{cases}
% \left(\frac{q_{t+1}(1)}{a_t(1)},\frac{q_{t+1}(2)}{a_t(2)}\right)^T-\pi_{\m{H}_1}(\vphi_t) &    \vphi_t\in \m{A}_1\\
% \left(\frac{q_{t+1}(1)}{a_t(1)},\frac{q_{t+1}(2)}{a_t(2)}\right)^T-\pi_{\m{H}_2}(\vphi_t) &    \vphi_t\in \m{A}_2\\
% \left(\frac{q_{t+1}(1)}{a_t(1)},\frac{q_{t+1}(2)}{a_t(2)}\right)^T-(1,1)^T & \vphi_t\in\m{A}_3\\
% (2q_{t+1}(1),2q_{t+1}(2))^T-\vphi_t & \vphi_t\in S
% \end{cases},
% \end{align}
% Consequently,
% \begin{align}
% \frac{(\vphi_t-\eta_t)^T (f(r_{t+1},q_{t+1})-\eta_t)}{\|\vphi_t-\eta_t\|_1}= a_t^T  \left(f(r_{t+1},q_{t+1})-\eta_t\right)=  \begin{cases}
% q_{t+1}(1)+q_{t+1}(2)-\lambda_1^T\pi_{\m{H}_1}(\vphi_t) &    \vphi_t\in \m{A}_1\\
% q_{t+1}(1)+q_{t+1}(2)-\lambda_2^T\pi_{\m{H}_2}(\vphi_t) &    \vphi_t\in \m{A}_2\\
% q_{t+1}(1)+q_{t+1}(2)-(a_t(1)+a_t(2)) & \vphi_t\in\m{A}_3\\
% 0 & \vphi_t\in S.
% \end{cases}
% \end{align}
% Recall that $q_{t+1}$ is in the simplex, so that $q_{t+1}(1)+q_{t+1}(2)=1$, and similarly $a_t(1)+a_t(2)=1$, by definition. Furthermore, we have that $\lambda_1^T \pi_{\m{H}_1}(\vphi_t)=1$ and $\lambda_2^T \pi_{\m{H}_2}(\vphi_t)=1$ by definition. Thus, $(\vphi_t-\eta_t)^T (f(r_{t+1},q_{t+1})-\eta_t)=0$ in all four regions $\m{A}_1,\m{A}_2,\m{A}_3,\m{S}$. 
Consequently,~\eqref{eq:rhoPartial} gives
\begin{align}
\rho_{t+1}\leq\frac{1}{(t+1)^2}\left(t^2 \rho_t+\|f(r_{t+1},q_{t+1})-\eta_t \|^2\right).\label{eq:rhotrecursion}
\end{align}
Telescoping~\eqref{eq:rhotrecursion}, we obtain that 
\begin{align}
\rho_t\leq\frac{1}{t^2}\sum_{i=1}^t \|f(r_i,q_i)-\eta_{i-1}\|^2,\label{eq:rhotelescoped}
\end{align}
where we defined $\eta_0=(0,0)^\tr$. We turn to develop an upper bound on $\|f(r_{t},q_{t})-\eta_{t-1} \|^2$ that holds for all $t$. We first write
\begin{align}
 \|f(r_{t},q_{t})-\eta_{t-1} \|^2\leq 2\|f(r_{t},q_{t})-\eta_{t-1} \|_{\infty}^2\leq 2\left(\|f(r_{t},q_{t})\|_{\infty}+\|\eta_{t-1}\|_\infty \right) ^2.
 \label{eq:l2linftybound}
\end{align}
Since $\eta_{t-1}\in\m{S}$, we have that $\|\eta_{t-1}\|_\infty\leq \Delta$.
Note that for any $q_t$ in the simplex
\begin{align}
\|f(r_t,q_t)\|_{\infty}\leq \max\left\{\frac{1}{r_t(1)},\frac{1}{r_{t}(2)} \right\}=\frac{1}{\min\{r_t(1),r_t(2)\}}.
\label{eq:lossinftynorm}
\end{align}
For $r_t=\lambda_1$, as well as $r_t=\lambda_2$ we have that $\min\{r_{t}(1),r_t(2)\}=\frac{\Delta-1}{\Delta}$ (recall that $\Delta\in(1,2]$).
For any $\bar{\vphi}_{t-1}\in\m{A}_3$, the vector $r_t=\frac{\bar{\vphi}_{t-1}-(1,1)^T}{\|\bar{\vphi}_{t-1}-(1,1)^T\|_1}$ is a convex combination of $\lambda_1$ and $\lambda_2$, and consequently $\min\{r_{t}(1),r_t(2)\}\geq\frac{\Delta-1}{\Delta}$ for all such vectors. For $r_t=(1/2,1/2)$ we clearly have $\min\{r_t(1),r_t(2)\}=\frac{1}{2}\geq \frac{\Delta-1}{\Delta}$.
Thus, by~\eqref{eq:lossinftynorm}, for any $r_t$ chosen by the algorithm~\eqref{eq:BlackwellUpdateRule}, and any $q_t$ in the simplex, it holds that
\begin{align}
\|f(r_t,q_t)\|_{\infty} \leq \frac{\Delta}{\Delta-1}.
\end{align}
Substituting this into~\eqref{eq:l2linftybound}, we obtain
\begin{align}
 \|f(r_{t},q_{t})-\eta_{t-1} \|^2\leq 2\left(\frac{\Delta}{\Delta-1}+\Delta \right)^2=\frac{2\Delta^4}{(\Delta-1)^2},
 \label{eq:l2linftybound2}
\end{align}
so that, by~\eqref{eq:rhotelescoped},
\begin{align}
\rho_t\leq \frac{2}{t}    \frac{\Delta^4}{(\Delta-1)^2}.
\end{align}
Finally,
\begin{align}
\|\bar{\vphi}_t\|_{\infty}\leq \|\bar{\vphi}_t-\eta_{t}\|_{\infty}+\|\eta_t\|_{\infty}\leq \sqrt{\rho_t}+\|\eta_t\|_{\infty}\leq \sqrt{\frac{2}{t}}\frac{\Delta^2}{\Delta-1}+\Delta.
\end{align}
\end{proof}

\section{Conclusions and Open Problems}
\label{sec:conc}

We have introduced the online bookmaking game, where the house/bookmaker updates the odds it offers based on the bets accumulated so far, with the goal of maximizing its worst-case return. We have shown that the optimal strategy of the gambler is decisive, meaning that in each round the gambler places all its bet on a single outcome. Consequently, the problem of designing a general optimal house strategy reduces to  designing an optimal strategy for decisive gamblers. For the binary case, we have shown that the latter problem is a special case of the tree bi-balancing problem, that we introduce and solve. Consequently, we obtain the complete solution to the binary online bookmaking problem, and develop the optimal algorithm for updating the odds offered by the house.

There are many follow-up questions arising from this work. The most obvious one is whether the optimal solution for the general online bookmaking game ($m>2$) can be obtained? To this end, one has to balance $m$-ary trees, $m>2$, and extending our techniques to handle the $m>2$ case seems to be a highly non-trivial task. Our algorithm is designed for a known horizon $T$, that is, the house knows how many rounds will be played, and the optimal strategy highly depends on $T$. Can we design a strategy that does not depend on the horizon $T$ and is near-optimal for all $T$? Note that the Blackwell-approachability-inspired algorithm we developed in Section~\ref{sec:Blackwell} does not depend on the horizon. However, it attains a loss of $T+\m{O}(T^{3/4})$ for all $T$, whereas the optimal loss is $L^*_T=T+T^{1/2}$. Finally, the tree bi-balancing technique we developed here may be applicable to other problems and loss functions. The key property our construction used is that the pair of values we can get for a tree of depth one is parametrized as $(x,f(x))$ where the function $f(x)$ is an involution. We believe that other important problems should have this property.

\bibliography{ref}
\bibliographystyle{IEEEtran}

\end{document}